\documentclass[aps,pre,twocolumn,superscriptaddress,showpacs,showkeys,amsmath,amssymb,footinbib,longbibliography]{revtex4-1}

\usepackage[T1]{fontenc}
\usepackage[utf8]{inputenc}
\usepackage{textcomp}
\usepackage{multirow}
\usepackage{float}
\usepackage{color}
\usepackage{bm}
\usepackage{amsmath}
\usepackage{amssymb}
\usepackage{graphicx}

\usepackage[breaklinks,colorlinks,bookmarks=false,citecolor=blue,linkcolor=red,urlcolor=blue]{hyperref}

\makeatletter

\usepackage{latexsym}
\usepackage[ngerman,english]{babel}

\newcommand{\braket}[2]{\langle \, {#1} \, | \, {#2} \, \rangle}

\def\ket#1{\, | \, {#1} \, \rangle}

\renewcommand{\theequation}{\arabic{section}.\arabic{equation}}

\begin{document}

\title{Adapting Planck's route to investigate the thermodynamics\\ 
       of the spin-half pyrochlore Heisenberg antiferromagnet}

\author{Oleg Derzhko}
\affiliation{Institute for Condensed Matter Physics,
          National Academy of Sciences of Ukraine,
          Svientsitskii Street 1, 79011 L'viv, Ukraine}
\affiliation{Max-Planck-Institut f\"{u}r Physik komplexer Systeme, 
          N\"{o}thnitzer Stra\ss e 38, 01187 Dresden, Germany}       
                    
\author{Taras Hutak}
\affiliation{Institute for Condensed Matter Physics,
          National Academy of Sciences of Ukraine,
          Svientsitskii Street 1, 79011 L'viv, Ukraine}
          
\author{Taras Krokhmalskii}
\affiliation{Institute for Condensed Matter Physics,
          National Academy of Sciences of Ukraine,
          Svientsitskii Street 1, 79011 L'viv, Ukraine}
          
\author{J\"{u}rgen Schnack}          
\affiliation{Fakult\"{a}t f\"{u}r Physik, Universit\"{a}t Bielefeld,
          Postfach 100131, 33501 Bielefeld, Germany}

\author{Johannes Richter}
\affiliation{Institut f\"{u}r Physik,
          Otto-von-Guericke-Universit\"{a}t Magdeburg,
          P.O. Box 4120, 39016 Magdeburg, Germany}
\affiliation{Max-Planck-Institut f\"{u}r Physik komplexer Systeme, 
          N\"{o}thnitzer Stra\ss e 38, 01187 Dresden, Germany}       

\date{\today}

\begin{abstract}
The spin-half pyrochlore Heisenberg antiferromagnet (PHAF) 
is one of the most challenging problems in the field of highly frustrated quantum magnetism. 
Stimulated by the seminal paper of M.~Planck 
[M.~Planck, Verhandl. Dtsch. phys. Ges. {\bf 2}, 202-204 (1900)] 
we calculate thermodynamic properties of this model by interpolating between the low- and high-temperature behavior.
For that we follow ideas developed in detail by B.~Bernu and G.~Misguich 
and use for the interpolation the entropy exploiting sum rules [the ``entropy method'' (EM)].
We complement the EM results for the specific heat, the entropy, and the susceptibility by corresponding results obtained 
by the finite-temperature Lanczos method (FTLM) for a finite lattice of $N=32$ sites 
as well as  
by the high-temperature expansion (HTE) data. 
We find that due to pronounced finite-size effects the FTLM data for $N=32$ are not representative for the infinite system below $T \approx 0.7$. 
A similar restriction to $T \gtrsim 0.7$ holds for the HTE designed for the infinite PHAF. 
By contrast,
the EM provides reliable data for the whole temperature region for the infinite PHAF.
We find evidence for a gapless spectrum leading to a power-law behavior of the specific heat at low $T$ 
and for a single maximum in $c(T)$ at $T\approx 0.25$. 
For the susceptibility $\chi(T)$ we find indications of a monotonous increase of $\chi$ upon decreasing of $T$ reaching $\chi_0 \approx 0.1$ at $T=0$.
Moreover, 
the EM allows to estimate the ground-state energy to $e_0\approx -0.52$.
\end{abstract}

\pacs{
75.10.-b, 
75.10.Jm  
}

\keywords{quantum Heisenberg antiferromagnet, 
pyrochlore lattice,  
finite-temperature Lanczos method,
high-temperature expansion, 
entropy interpolation method}

\maketitle

\section{Introduction}
\label{sec1}
\setcounter{equation}{0}

A paradigmatic highly frustrated spin model is the pyrochlore Heisenberg antiferromagnet (PHAF).
The pyrochlore lattice is built of corner-sharing tetrahedra, 
see Fig.~\ref{fig01}, below. 
There are several compounds where the magnetic atoms reside on the sites of the pyrochlore lattice and the exchange interaction is antiferromagnetic,
see, e.g., Refs.~\cite{Gardner_2010,Gingras_McClarty_2014,Rau2018}.

Already the classical PHAF
(i.e., for spin $S\to\infty$) 
exhibits interesting properties and its study is far from being trivial
\cite{Reimers1991,Reimers1992,Moessner1998a,Moessner1998b,Isakov_2004,Henley_2010,Lapa2012}.
Thus, 
the ground-state manifold is highly degenerate, 
the model exhibits strong short-range correlations, but it does not exhibit any long-range order,
and, 
because of the huge degeneracy of the ground state, 
the model is very susceptible to various perturbations.

The quantum spin $S=1/2$ PHAF is even more complicated.
Thus, so far no accurate values for the ground-state energy $e_0$ for this model are available. 
On the one hand, 
the $S=1/2$ case opens the route to new quantum phases \cite{FPRG_Pyro_2018}.
On the other hand,
such powerful straightforward numerical tools like standard quantum Monte Carlo or molecular dynamics simulations are not applicable for the $S=1/2$ PHAF.
Moreover, several approximation methods developed for one- and two-dimensional quantum spin systems
(e.g., density matrix renormalization group and tensor network methods) 
are very limited in three dimensions.

Theoretical studies of the quantum PHAF are mostly focused on ground-state properties, 
see, e.g., 
\cite{Harris1991,Isoda1998,Canals1998,Canals2000,Koga2001,Tsunetsugu2001,Tsunetsugu2001b,
Berg2003,Moessner_2006,Tchernyshyov_2006,Kim2008,Burnell2009,Chandra_ED_pyro_2018,FPRG_Pyro_2018,RGM_pyro_2019},
whereas much less attention has been paid to its finite-temperature properties.
One reason for that is the lack of methods to study thermodynamics of three-dimensional frustrated quantum spin systems.
Among the few papers studying the thermodynamics of the $S=1/2$ PHAF
we mention bold diagrammatic Monte Carlo simulations 
(stochastic sampling of all skeleton Feynman diagrams) 
down to the temperature $J/6$ \cite{Huang2016}.
This paper reports data for the susceptibility $\chi(T)$ but no data for the specific heat $c(T)$.
We will refer to these data for $\chi(T)$ in Sec.~\ref{sec4B}.
A comprehensive analysis of the spin-$S$ $J_1-J_2$ Heisenberg model
by employing the pseudofermion functional renormalization group technique was presented in Ref.~\cite{FPRG_Pyro_2018}. 
However, this paper does not contain data for $\chi(T)$ and $c(T)$.
Finally, 
we mention 
the high-temperature expansion study 
and 
the rotation-invariant Green's function study 
of the $S=1/2$ PHAF \cite{Lohmann2011,RGM_pyro_2019}.
In these recent papers \cite{Huang2016,FPRG_Pyro_2018,RGM_pyro_2019} 
no evidence for a finite-temperature phase transition was found, 
i.e., the spin-half PHAF is most likely a three-dimensional spin system 
without singularities in the specific heat and the susceptibility.  

The goal of the present paper is to study the thermodynamics of the $S=1/2$ PHAF for the whole temperature region 
focussing on the specific heat $c(T)$ and the static uniform susceptibility $\chi(T)$, 
both being basic and easily accessible quantities in experimental studies of PHAF compounds.
To deal with the above mentioned challenges when studying the finite-temperature properties of the $S=1/2$ PHAF, 
we follow M.~Planck's ideas of his seminal paper in 1900 \cite{Planck1}, see also Appendix~A,  
and perform a sophisticated interpolation between the low- and high-temperature behavior of a thermodynamic potential, 
namely, the entropy $s$ as a function of internal energy $e$. 
For that we exploit also sum rules valid for the specific heat as proposed by B.~Bernu and G.~Misguich \cite{Bernu2001,Misguich2005}, 
for details see Sec.~\ref{sec3A}.
In what follows we call this approach the entropy method (EM).
We complement our studies based on the EM 
by using 
the finite-temperature Lanczos method (FTLM) for a finite pyrochlore lattice of $N=32$ sites 
and  
the high-temperature expansion (HTE) up to order 13.

In the present paper, 
we estimate the ground-state energy to $e_0\approx -0.52$
and find evidence for a gapless spectrum, i.e., for a power-law behavior of the specific heat at low temperatures, 
and for a single maximum in $c(T)$ at about 25\% of the exchange coupling.

The manuscript is organized as follows.
We begin with introducing the model (Sec.~\ref{sec2}) and the description of the exploited methods (Sec.~\ref{sec3}).
We report our findings obtained by the FTLM (finite lattices) and by the HTE and EM (infinite lattice) in Sec.~\ref{sec4}.
We summarize our results in Sec.~\ref{sec5}.

\section{Model}
\label{sec2}
\setcounter{equation}{0}

\begin{figure}
\begin{center}
\includegraphics[clip=on,width=80mm,angle=0]{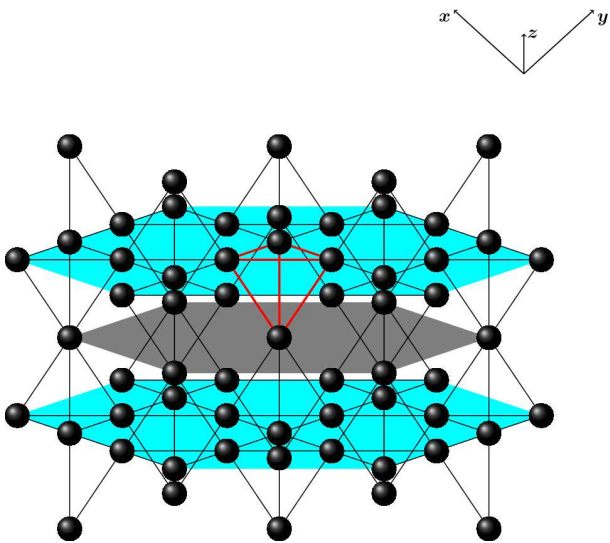}
\caption{The pyrochlore lattice 
visualized here as a three-dimensional structure which consists of alternating kagome (cyan) and triangular (gray) planar layers. 
The four-site unit cell is marked with the red bonds.}
\label{fig01} 
\end{center}
\end{figure}

We consider the Heisenberg model on the pyrochlore lattice (see Fig.~\ref{fig01}) given by the Hamiltonian
\begin{eqnarray}
\label{201}
H = \sum_{\langle m\alpha,n\beta\rangle} {\bm{S}}_{m\alpha}\cdot {\bm{S}}_{n\beta}.
\end{eqnarray}
We have set the antiferromagnetic nearest-neighbor coupling to unity, $J=1$, fixing the energy scale.
The sum in Eq.~(\ref{201}) runs over all nearest-neighbor bonds and $\boldsymbol{S}_{m\alpha}^2=3/4$.

The pyrochlore lattice consists of four interpenetrating face-centered-cubic sublattices.
The origins of these four sublattices are located at 
${\bf{r}}_1=(0,0,0)$,
${\bf{r}}_2=(0,1/4,1/4)$,
${\bf{r}}_3=(1/4,0,1/4)$,
and
${\bf{r}}_4=(1/4,1/4,0)$.
The sites of the face-centered-cubic lattice are determined by 
${\bf{R}}_m=m_1{\bf{e}}_1+m_2{\bf{e}}_2+m_3{\bf{e}}_3$,
where $m_1$, $m_2$, $m_3$ are integers
and
${\bf{e}}_1=(0,1/2,1/2)$,
${\bf{e}}_2=(1/2,0,1/2)$,
${\bf{e}}_3=(1/2,1/2,0)$.
As a result,
the $N$ pyrochlore lattice sites are labeled by $m\alpha$,
${\bf{R}}_{m\alpha}={\bf{R}}_m+{\bf{r}}_\alpha$,
where $m=1,\ldots, {\cal{N}}$, 
${\cal{N}}=N/4$ is the number of unit cells,
and $\alpha=1,2,3,4$ labels the sites in the unit cell.

There are a few compounds 
with magnetic atoms residing on pyrochlore-lattice sites        
with antiferromagnetic nearest-neighbor exchange interactions,
which can be considered as experimental realizations of the quantum PHAF. 
Besides the fluoride NaCaNi$_2$F$_7$ which provides a good realization of the $S=1$ PHAF \cite{Plumb2017,Zhang2018},
we may mention 
the molybdate Y$_2$Mo$_2$O$_7$ \cite{Greedan1986,Silverstein2014,Thygesen2017},
the chromites $A$Cr$_2$O$_4$ ($A$=Mg,Zn,Cd) \cite{Gao2018,Ji2009,Matsuda2007}, 
or
FeF$_3$ \cite{Sadeghi2015}.
Unfortunately,
we are not aware of any solid-state realization of the PHAF model with $S=1/2$ given in Eq.~(\ref{201}).

\section{Methods}
\label{sec3}
\setcounter{equation}{0}

\subsection{Entropy method (EM)}
\label{sec3A}

In accordance with M.~Planck's strategy to derive the energy distribution of the black-body radiation \cite{Planck1,Planck2}, 
the EM is an interpolation scheme that combines presumed knowledge on high- and low-temperature properties
and, in addition, exploits sum rules for the specific heat $c(T)$ in a clever way.
The EM as used in the present paper was introduced in 2001 by B.~Bernu and G.~Misguich \cite{Bernu2001}. 
The method has been afterwards used, modified, and extended in Refs.~\cite{Misguich2005,Bernu2013,Bernu2015,Schmidt_2017,Bernu2019}.
Below we explain briefly this procedure for self consistency.

Within the framework of the EM,
we use the microcanonical ensemble working with the entropy per site $s$ as a function of the energy per site $e$, 
$s(e)$,
in the whole (finite) range of energies.
The temperature $T$ and the specific heat per site $c$ are given by the formulas
\begin{eqnarray}
\label{301}
T=\frac{1}{s^\prime},
\;\;\;
c=-\frac{{s^\prime}^2}{s^{\prime\prime}},
\end{eqnarray}
where the prime denotes the derivative with respect to $e$.
These equations form a parametric representation of the dependence $c(T)$.
Knowing the high-temperature series for $c(T)$ up to $n$th order, 
$c(T)=\sum_{i=2}^nd_i\beta^i+{\cal {O}}(\beta^{n+1})$ ($d_1=0$), $\beta=1/T$,
we immediately get the series for $s(e)$ around the maximal energy $e_\infty=0$ up to the same order $n$,
\begin{eqnarray}
\label{302}
\left.s(e)\right\vert_{e\to e_\infty=0}\to \ln 2+\sum_{i=2}^na_ie^i,
\end{eqnarray}
where the coefficients $a_i$ are known functions of the coefficients $d_i$, see Appendix~A of Ref.~\cite{Bernu2001}.
The behavior of $s(e)$ as $e$ approaches the (minimal) ground-state energy $e_0$ 
(i.e., as the temperature approaches 0)  
is also supposed to be known.
It is,
\begin{eqnarray}
\label{303}
\left.s(e)\right\vert_{e\to e_0}
\propto 
\left(e-e_0\right)^{\frac{\alpha}{1+\alpha}}
\end{eqnarray}
if $c(T)$ vanishes as $T^\alpha$ when $T\to 0$ (gapless excitations)
and
\begin{eqnarray}
\label{304}
\left.s(e)\right\vert_{e\to e_0}
\propto 
-\frac{e-e_0}{\Delta}\left(\ln\left[\Delta\left(e-e_0\right)\right]-1\right)
\end{eqnarray}
if $c(T)$ vanishes as $T^{-\alpha}\exp(-\Delta/T)$, $\alpha=2$, when $T\to 0$ (gapped excitations).
Therefore we proceed differently in the gapless case and in the gapped case.
Here it is assumed that $e_0$ and $\alpha$ are known (gapless case) or $e_0$ is known and $\alpha=2$ (gapped case).

In the gapless case,
we introduce the auxiliary function \cite{Misguich2005}
\begin{eqnarray}
\label{305}
G(e)=\frac{\left(s(e)\right)^{\frac{1+\alpha}{\alpha}}}{e-e_0} 
\end{eqnarray}
and approximate it as
\begin{eqnarray}
\label{306}
G_{\rm app}(e)=G(0)[u,d](e),
\;\;\;
G(0)=\frac{\left(\ln 2\right)^{\frac{1+\alpha}{\alpha}}}{-e_0}.
\end{eqnarray}
Here $[u,d](e)=P_u(e)/Q_d(e)$ is a Pad\'{e} approximant, 
where the coefficients of the polynomials $P_u(e)$ and $Q_d(e)$ 
(of order $u$ and $d$, respectively, $u+d\le n$)
are determined by the condition that the expansion of $[u,d](e)$
has to agree with the power series of $G(e)/G(0)$ [which follows from Eqs.~(\ref{305}) and (\ref{302})] up to order ${\cal {O}}(e^{u+d})$.
Of course, $G(0)=G_{\rm app}(0)$.
The approximate entropy follows by inverting Eq.~(\ref{305})
\begin{eqnarray}
\label{307}
s_{\rm app}(e)=\left[\left(e-e_0\right)G_{\rm app}(e)\right]^{\frac{\alpha}{1+\alpha}}.
\end{eqnarray}
The prefactor $A$ in the power-law decay of the specific heat $c(T)$ for $T\to 0$,
$c(T)\to AT^\alpha$,
is given by
\begin{eqnarray}
\label{308}
A_{\rm app}=\frac{\alpha^{1+\alpha}}{\left(1+\alpha\right)^\alpha}\left[G_{\rm app}(e_0)\right]^\alpha.
\end{eqnarray}

In the gapped case,
we introduce the auxiliary function \cite{Bernu2001}
\begin{eqnarray}
\label{309}
G(e)=\left(e-e_0\right)\left(\frac{s(e)}{e-e_0}\right)^\prime
\end{eqnarray}
and approximate it as
\begin{eqnarray}
\label{310}
G_{\rm app}(e)=G(0)[u,d](e),
\;\;\;
G(0)=\frac{\ln 2}{e_0}.
\end{eqnarray} 
Here $[u,d](e)=P_u(e)/Q_d(e)$ again is a Pad\'{e} approximant, 
where the coefficients of the polynomials $P_u(e)$ and $Q_d(e)$ 
(of order $u$ and $d$, respectively, $u+d\le n$)
are determined by the condition that the expansion of $[u,d](e)$
has to agree with the power series of $G(e)/G(0)$ [which follows now from Eqs.~(\ref{309}) and (\ref{302})] up to order ${\cal {O}}(e^{u+d})$.
Of course, $G(0)=G_{\rm app}(0)$.
The approximate entropy follows by inverting Eq.~(\ref{309})
\begin{eqnarray}
\label{311}
\frac{s_{\rm app}(e)}{e-e_0}
=
\frac{\ln 2}{-e_0}-\int\limits_{e_0\le e\le 0}^0{\rm d}\xi\frac{G_{\rm app}(\xi)}{\xi - e_0}.
\end{eqnarray}
From the technical point of view, 
before performing the integration in the right-hand side of Eq.~(\ref{311})
one may perform the partial fraction expansion of the integrand which is obviously a rational function.
The excitation gap $\Delta$ in the decay of the specific heat $c(T)$ for $T\to 0$, 
$c(T)\propto T^{-2}\exp(-\Delta/T)$,
is given by
\begin{eqnarray}
\label{312}
\Delta_{\rm app}=-\frac{1}{G_{\rm app}(e_0)}.
\end{eqnarray}

Until now we considered the EM for zero magnetic field $h=0$.
Of course, for non-zero $h$ the thermodynamic functions depend on $h$, 
i.e., the entropy is now $s(e,h)$. 
The magnetization per site $m$ and the uniform susceptibility per site $\chi$ are given by the formulas \cite{Bernu2019}
\begin{eqnarray}
\label{313}
m=\frac{1}{(s(e,h))^\prime}\frac{\partial s(e,h)}{\partial h},
\;\;\;
\chi=\frac{m}{h} ,
\end{eqnarray}
where the last equation implies that $h$ is infinitesimally small.
Clearly, the HTE coefficients for the specific heat are also changed. 
Simple algebra yields
\begin{eqnarray}
\label{314}
d_i\rightarrow d_i+\frac{(i-1)i}{2}c_{i-1}h^2,
\;\;\; 
i=2,\ldots,n;
\end{eqnarray}
we use here the high-temperature series for the static uniform susceptibility $\chi(T)=\sum_{i=1}^{n}c_i\beta^i+{\cal {O}}(\beta^{n+1})$, $\beta=1/T$.
The expression (\ref{302}) for the series of $s$ is valid, 
however, the coefficients $a_i$ are now known functions of the coefficients $d_i$, $c_i$, and $h$. 
For the gapless case all reasonings in Eqs.~(\ref{303}), (\ref{305}) to (\ref{308}) hold 
with the only difference that the ground-state energy now is $e_0-\chi_0 h^2/2$, 
where $\chi_0\equiv\chi(T=0)$ is the ground-state susceptibility which is assumed to be known.
The approximate entropy in Eq.~(\ref{307}) now also depends on $h$, i.e., $s_{\rm app}(e,h)$.
For the case of gapped magnetic excitations the ground-state energy remains unchanged, 
because $\chi_0=0$,
and therefore all the equations (\ref{304}), (\ref{309}) to (\ref{312}) are valid.
Again, the approximate entropy in Eq.~(\ref{311}) now also depends on $h$, 
i.e., $s_{\rm app}(e,h)$.

In summary,
knowing the high-temperature series of $c(T)$ and $\chi(T)$ together with 
(i) the ground-state energy $e_0$, the exponent $\alpha$, and the ground-state susceptibility $\chi_0$
for the gapless case
or 
(ii) only the  ground-state energy $e_0$ 
for the gapped case,
we obtain $c(T)$ and $\chi(T)$ at all temperatures.
For that, we use $s_{\rm app}(e,h)$ 
which yields the specific heat $c(T)$ by Eq.~(\ref{301}) and the susceptibility $\chi(T)$ by Eqs.~(\ref{313}) and (\ref{301}).

Based on previous experience with the EM \cite{Bernu2001,Misguich2005,Bernu2013,Bernu2015,Schmidt_2017,Bernu2019},
we use the following strategy: 
We discard those Pad\'{e} approximants in $G_{\rm app}(e)$, Eqs.~(\ref{306}) and (\ref{310}),
which give unphysical solutions;
the remaining ones are called ``physical''.
Moreover, we focus on those input parameter sets 
for which interpolations based on different Pad\'{e} approximants 
lead to data sets for $c(T)$ and $\chi(T)$ being quite close to each other.
For further details about the EM in the context of the $S=1/2$ PHAF see Sec.~\ref{sec4B}.

\subsection{Finite-temperature Lanczos method (FTLM)}
\label{sec3B}

The FTLM is an efficient and very accurate approximation 
to calculate thermodynamic quantities of quantum spin systems on finite lattices of $N$ sites at arbitrary temperatures.
It is an unbiased numerical approach, 
where thermodynamic quantities such as the specific heat and the susceptibility are determined using trace estimators
\cite{JaP:PRB94,JaP:AP00,ScW:EPJB10,PrB:SSSSS13,HaS:EPJB14,ScT:PR17,PRE:COR17,kago42,FTLM_accuracy}.
The key element is the approximation of the partition function $Z$ using a Monte-Carlo like representation of $Z$, 
i.e., the sum over a complete set of $2^N$ basis vectors present in $Z$
is replaced by a much smaller sum over $R$ random vectors $\ket{\nu}$ for each subspace ${\mathcal H}(\gamma)$ of the Hilbert space, 
where except the conservation of total $S^z$ we also use the lattice symmetries of the Hamiltonian 
to decompose the full Hilbert space into mutually orthogonal subspaces labeled by $\gamma$. 
The exponential of the Hamiltonian is then approximated by its spectral representation in a Krylov space 
spanned by the $N_L$ Lanczos vectors starting from the respective random vector $\ket{\nu}$. 
The FTLM representation of the  partition function finally reads 
\begin{eqnarray}
\label{315}
Z(T)
&\approx&
\sum_{\gamma=1}^\Gamma\!
\frac{\text{dim}({\mathcal H}(\gamma))}{R}
\sum_{\nu=1}^R\!
\sum_{n=1}^{N_L}\!
\exp\left(\!-\frac{\epsilon_n^{(\nu)}}{T}\!\right)\!\vert\braket{n(\nu)}{\nu}\vert^2 ,
\nonumber\\[-3mm]
\end{eqnarray}
where $\ket{n(\nu)}$ is the $n$th eigenvector of $H$ in the Krylov space with the corresponding energy $\epsilon_n^{(\nu)}$.
To perform the symmetry-decomposed numerical Lanczos calculations   
we use J.~Schulenburg's {\it spinpack} code \cite{spin:256,RiS:EPJB10}.

\subsection{High-temperature expansion (HTE)}
\label{sec3C}

The HTE is a universal approach to discuss the thermodynamics of spin systems \cite{Oitmaa2006}.
In the present study we use the Magdeburg HTE code developed mainly by A.~Lohmann \cite{Lohmann2011,Lohmann-Diplom,Lohmann2014}
(which is freely available at \verb"http://www.uni-magdeburg.de/jschulen/HTE/")
in an extended version up to 13th order,
see Appendix~B.
With this tool,
we compute the series of the specific heat 
$c(T)=\sum_{i=2}^nd_i\beta^i+{\cal {O}}(\beta^{n+1})$ ($d_1=0$)
and the static uniform susceptibility 
$\chi(T)=\sum_{i=1}^{n}c_i\beta^i+{\cal {O}}(\beta^{n+1})$
with respect to the inverse temperature $\beta=1/T$.

To extend the region of validity of the ``raw'' HTE series we may use Pad\'{e} approximants $[m,n]=P_m(\beta)/Q_n(\beta)$,
where $P_m(\beta)$ and $Q_n(\beta)$ are polynomials in $\beta$ of order $m$ and $n$, respectively.
The coefficients of the polynomials $P_m(\beta)$ and $Q_n(\beta)$ are determined by the condition 
that the expansion of $[m,n]$ has to agree with the initial power series up to order ${\cal{O}}(\beta^{m+n})$.

\section{Results}
\label{sec4}
\setcounter{equation}{0}

\subsection{Finite lattices} 
\label{sec4A}

In Ref.~\cite{Chandra_ED_pyro_2018} finite lattices of $N=28$ and $N=36$ sites are used to discuss ground-state properties. 
These lattices are built by stacked alternating triangular and kagome layers 
imposing periodic boundary conditions within the layers, but open boundary conditions perpendicular to them. 
We have calculated the HTE series for these finite lattices.
We compare these finite-lattice series with the corresponding HTE series of the infinite pyrochlore lattice to judge the finite lattices. 
We found that for $c(T)$ all HTE coefficients are different. 
For the susceptibility $\chi(T)$ only the lowest-oder term coincides, i.e., the agreement is only marginally better.
This drastic difference between the finite lattices and the infinite pyrochlore lattice 
can be attributed to the edge spins stemming from the imposed open boundary conditions. 
Thus, we conclude that the finite lattices of $N=28$ and $N=36$ used in Ref.~\cite{Chandra_ED_pyro_2018} 
are not appropriate to discuss the thermodynamics of the PHAF.  
However, we note that they can be useful to discuss the ground-state properties, 
e.g., spin-spin correlations when considering spins away from the edge spins.

A more suitable finite lattice is the one with $N=32$ sites imposing periodic boundary conditions in all directions. 
This lattice contains eight face-centered-cubic cells, 
i.e., the edge vectors go along the face-centered-cubic basis vectors and have twice the length of these. 
For this lattice the HTE series for $c$ ($\chi$) coincides up to 3rd (4th) order with that of the infinite lattice.

\begin{figure}
\centering 
\includegraphics[clip=on,width=80mm,angle=0]{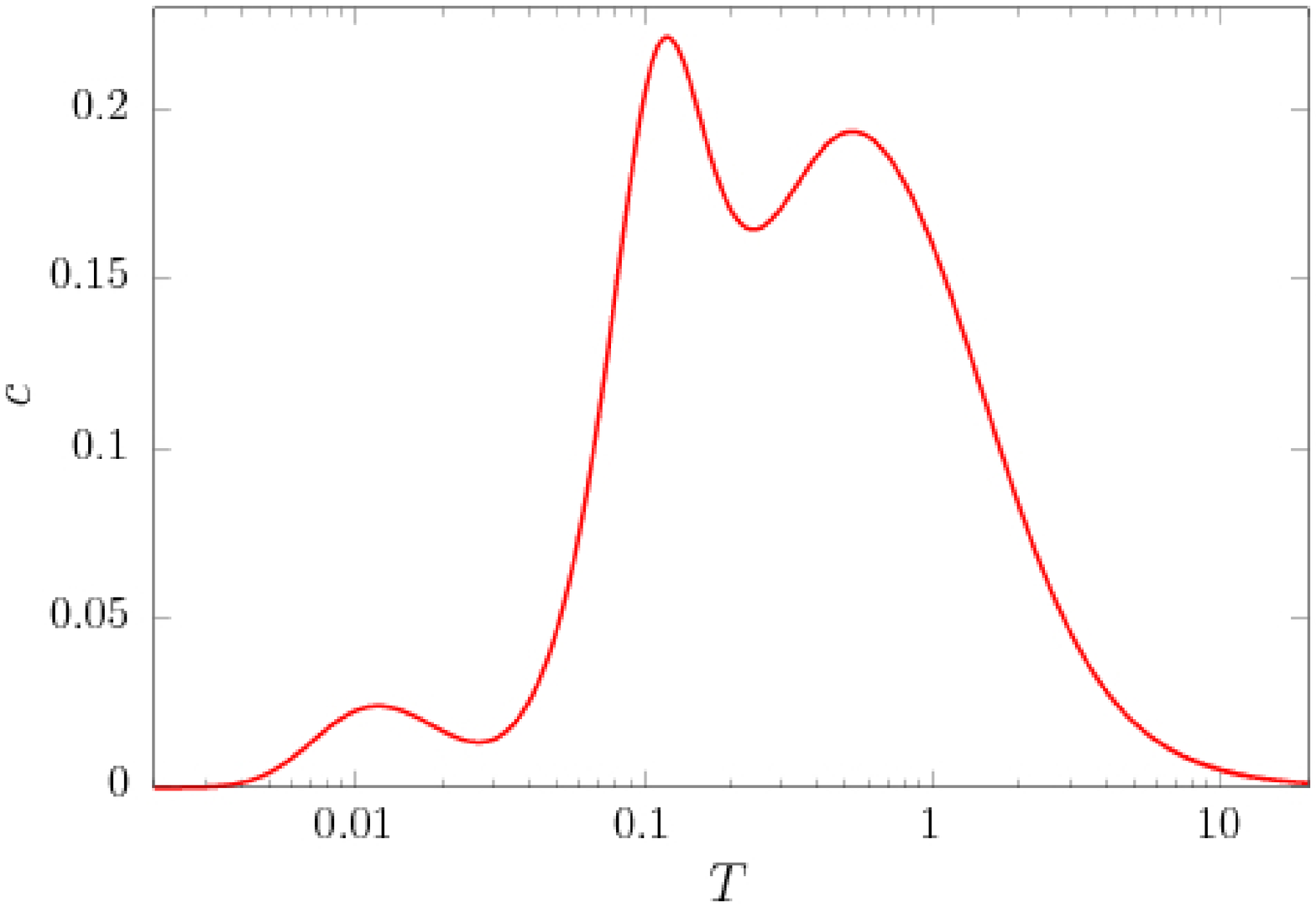}
\includegraphics[clip=on,width=80mm,angle=0]{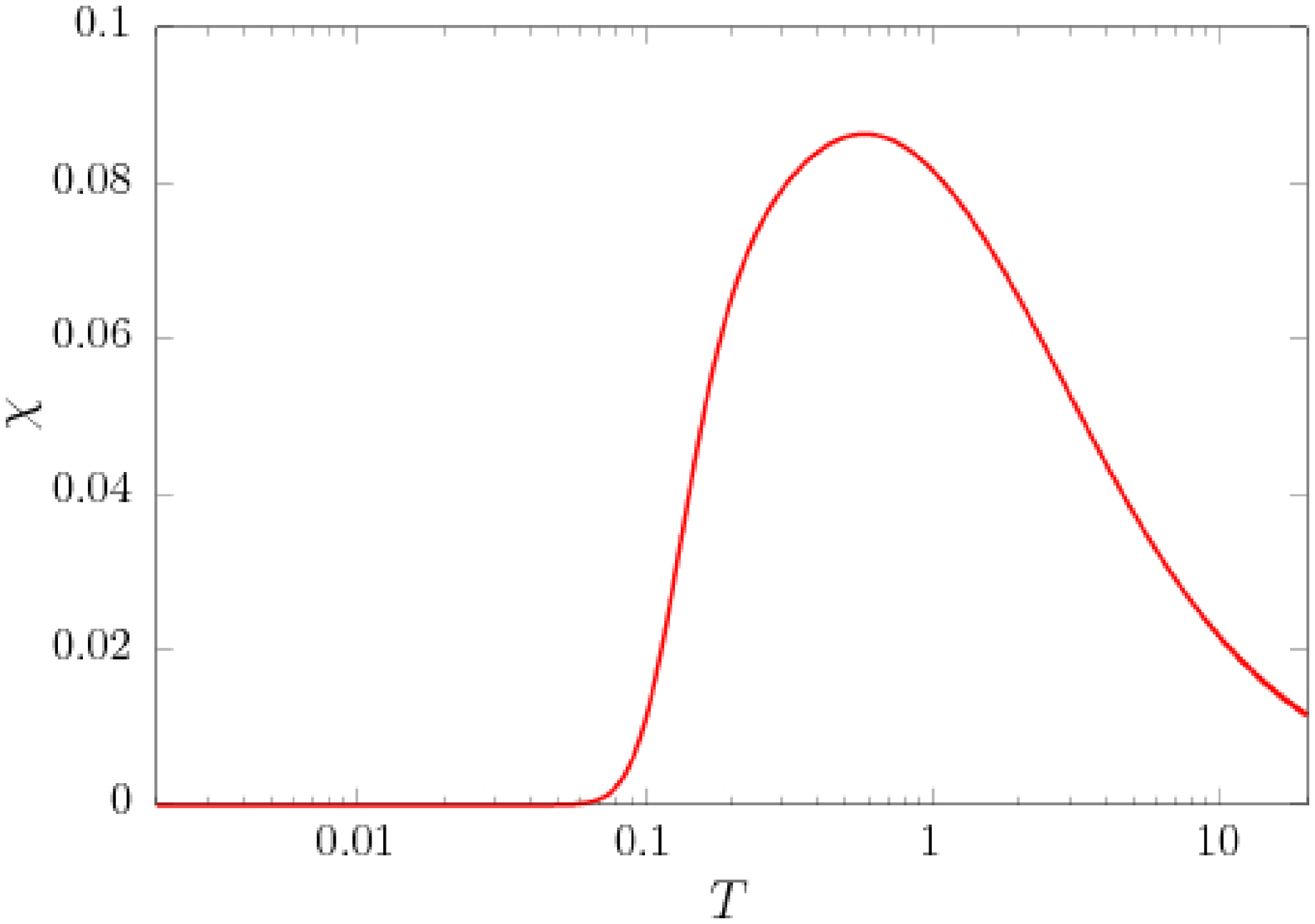}
\protect
\caption{FTLM data ($R=100$) for the temperature dependence (logarithmic scale) of 
(top) the specific heat per site $c(T)$
and 
(bottom) the static uniform susceptibility per site $\chi(T)$ 
of the PHAF of $N=32$ sites.} 
\label{fig02} 
\end{figure}

\begin{figure}[H]
\centering 
\includegraphics[clip=on,width=80mm,angle=0]{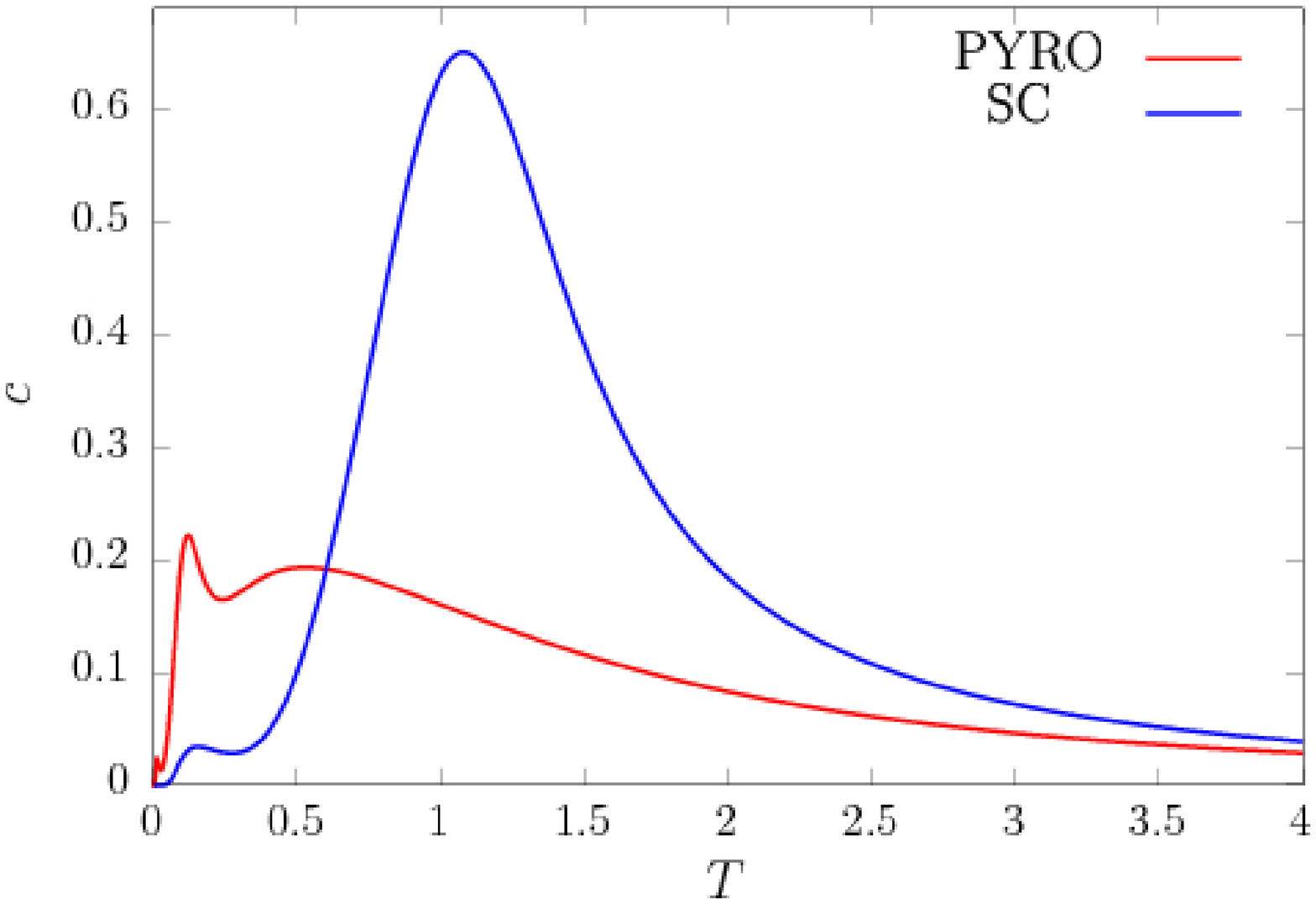}
\includegraphics[clip=on,width=80mm,angle=0]{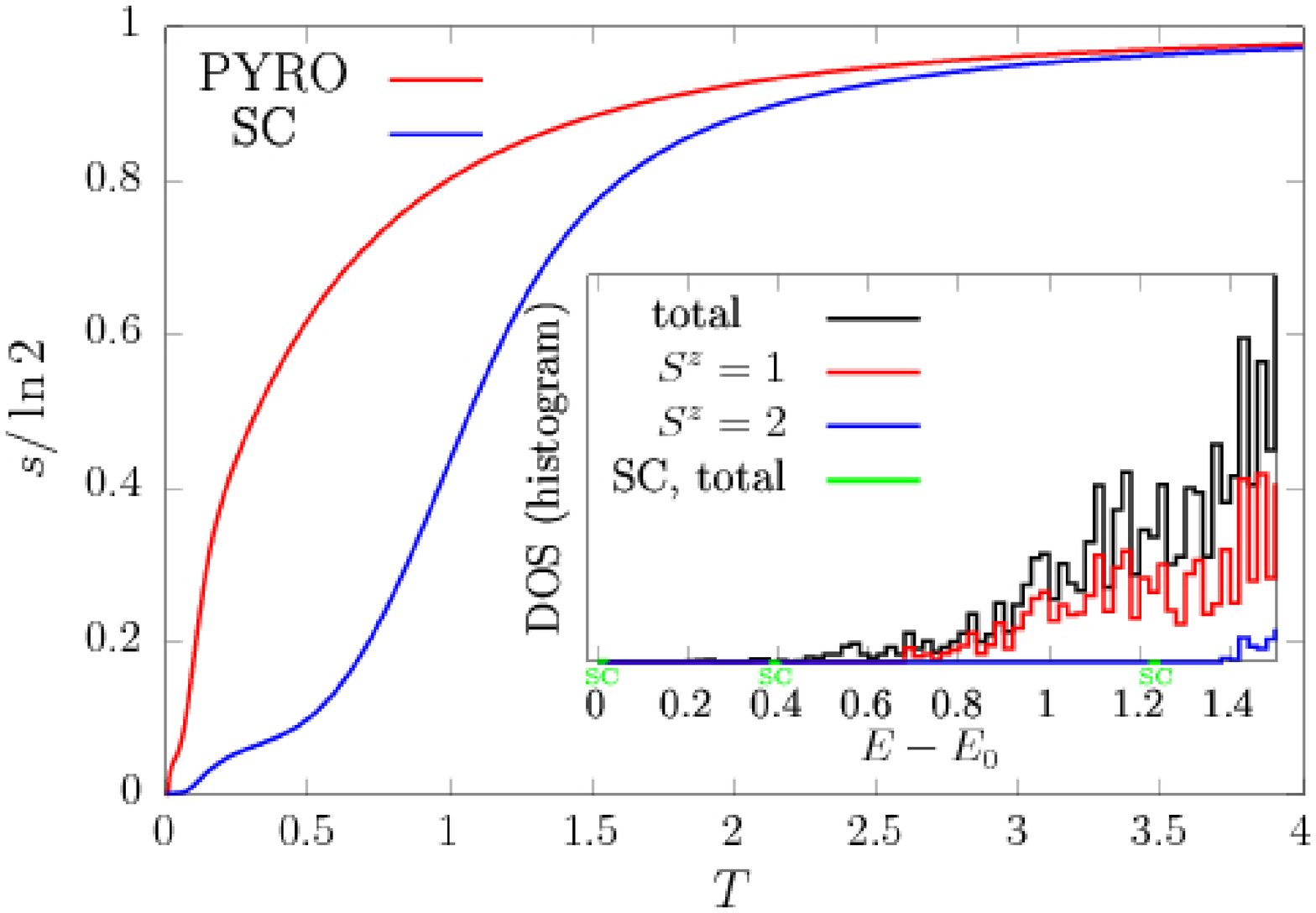}
\includegraphics[clip=on,width=80mm,angle=0]{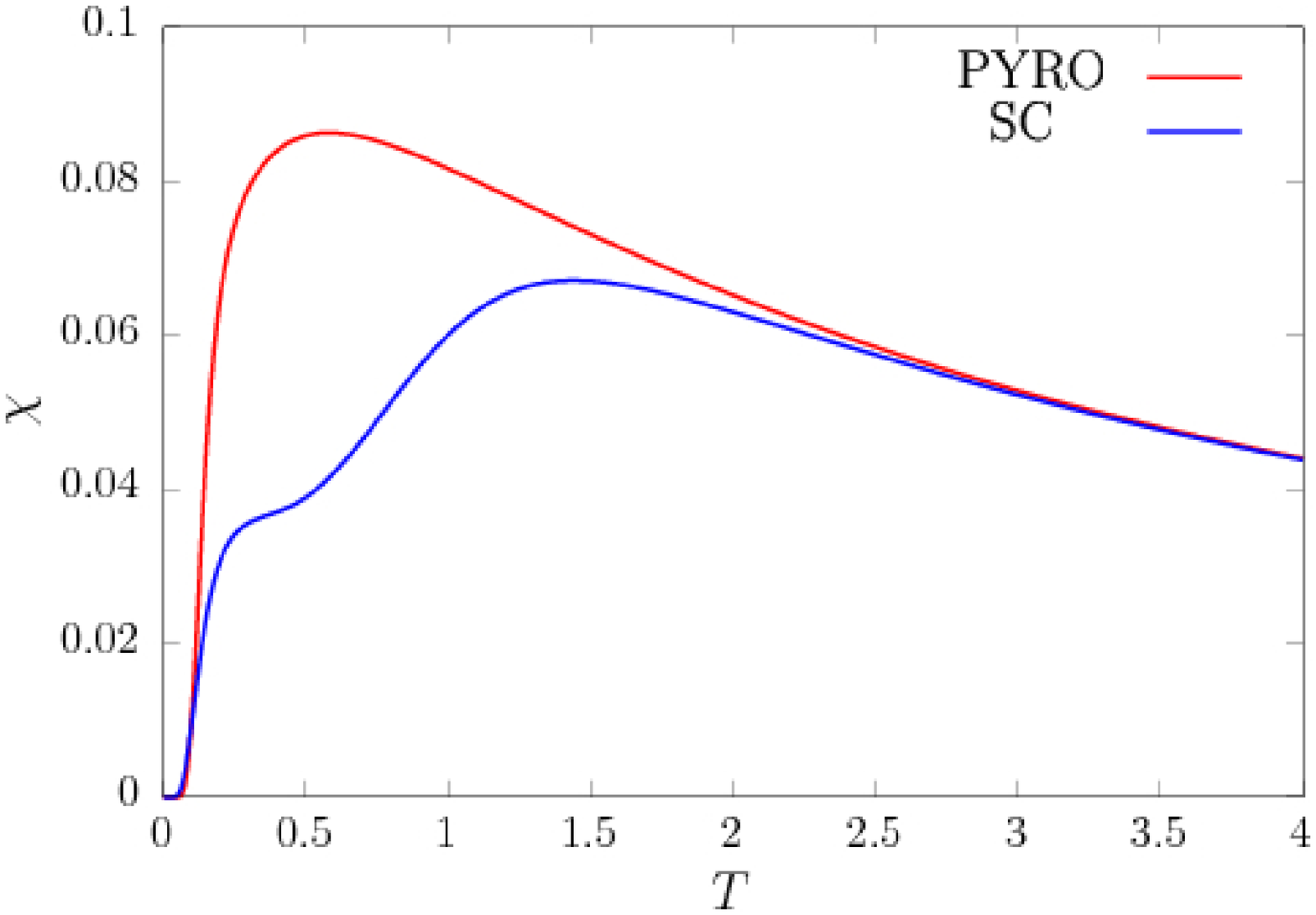}
\protect
\caption{Comparison of FTLM data ($R=100$) of the PHAF of $N=32$ sites with corresponding ones of the simple-cubic HAF of $N=32$ sites.
(Top) Specific heat per site $c(T)$.
(Middle) Entropy per site $s(T)$. 
(Bottom) Static uniform susceptibility per site $\chi(T)$.
The inset in the middle panel shows the histogram low-energy density of states (arbitrary units). 
Note that for the simple-cubic HAF there are only three tiny bars in the energy region shown here 
(their positions are indicated by green labels ``sc'').}
\label{fig03}
\end{figure}

The FTLM is the adequate approach to study the finite $S=1/2$ PHAF of $N=32$ sites.
In Fig.~\ref{fig02} we show data for $c(T)$ (top) and $\chi(T)$ (bottom) over a wide temperature range using a logarithmic $T$ scale.    
The specific heat exhibits the typical main maximum at $T=0.53$
and, in addition, two low-$T$ maxima at $T=0.012$ and at $T=0.117$. 
While the maximum at $T=0.012$ is certainly a finite-size effect, 
one can speculate that the other low-$T$ maximum at $T=0.117$ signals an extra low-energy scale set by low-lying singlets 
(see the density of states shown in the inset of the middle panel of Fig.~\ref{fig03}) 
that might be also relevant for the infinite system.
Such a feature has been observed in low-dimensional highly frustrated quantum magnets, 
e.g., the spin-half kagome Heisenberg antiferromagnet (HAF), 
where the existence of such an extra low-$T$ peak is a subject of a long-standing and ongoing debate 
\cite{elstner1994,NaM:PRB95,ToR:PRB96,Misguich2005,Mun:WJCMP14,Shimokawa2016,CRL:SB18,kago42}.
However, in the three-dimensional PHAF the finite-size effects are undoubtedly stronger than in the two-dimensional kagome HAF.
Thus, to conclude a double-peak structure in $c(T)$ from our FTLM data is inappropriate.
For the static uniform susceptibility $\chi(T)$ the low-lying singlets are not relevant 
and $\chi(T)$ does not show extra-peaks except the well-pronounced maximum
that is typical for finite spin systems with $\chi_0\equiv\chi(T=0)=0$.
Again, this behavior might be not representative for the infinite system,
particularly, in case that $\chi_0>0$ for $N\to\infty$.

\begin{figure}
\centering 
\includegraphics[clip=on,width=80mm,angle=0]{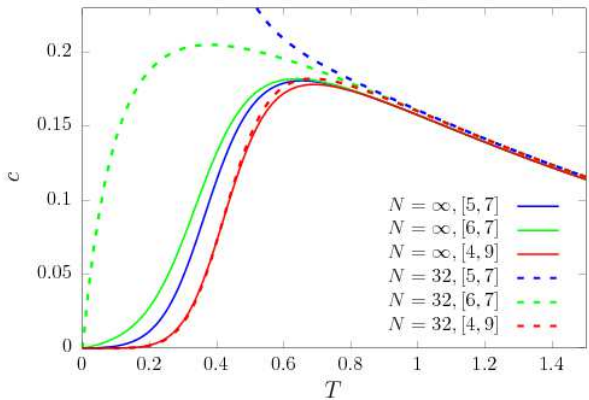}
\protect
\caption{Several Pad\'{e} approximants of the specific heat $c(T)$ of the PHAF: 
Comparison of the finite lattice of $N=32$ (broken) with the infinite lattice (solid).}
\label{fig04} 
\end{figure}

\begin{figure}
\centering 
\includegraphics[clip=on,width=80mm,angle=0]{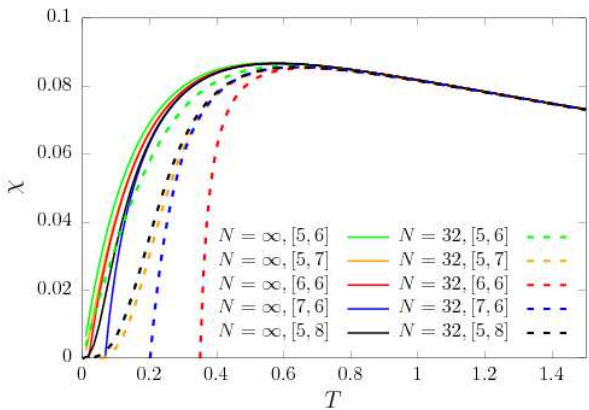}
\protect
\caption{Several Pad\'{e} approximants of the uniform susceptibility $\chi(T)$ of the PHAF: 
Comparison of the finite lattice of $N=32$ (broken) with the infinite lattice (solid).}
\label{fig05}
\end{figure}

To quantify the temperature region where the finite $N=32$ lattice may be representative for the infinite lattice 
we compare in Figs.~\ref{fig04} and \ref{fig05} several Pad\'{e} approximants of the HTE series of the finite and the infinite lattices. 
Obviously, the data for $N=32$ and $N=\infty$ coincide only down to $T \sim 0.7$, 
thus, indicating that finite pyrochlore lattices accessible by FTLM 
are not suitable to discuss the thermodynamics of the spin-half PHAF below this temperature.

Nevertheless, the finite-size data for the PHAF are useful to demonstrate frustration effects. 
For that we may compare the PHAF with the $S=1/2$ HAF on the simple-cubic lattice, 
because both have six nearest-neighbors,  
i.e., a simple mean-field decoupling of the Heisenberg Hamiltonian would yield identical thermodynamics.
However, the simple-cubic HAF exhibits a finite-temperature phase transition to N\'{e}el order, but the PHAF does not order. 
Thus, we compare both models on finite lattices of $N=32$ sites, 
where the simple-cubic finite-temperature phase transition is irrelevant,
see Fig.~\ref{fig03}, 
where we compare FTLM data of the specific heat, the entropy, and the susceptibility using a linear $T$ scale for $N=32$.
The tremendous influence of frustration is visible at all temperature scales.
In particular, the spectrum at low energies in the frustrated system is much denser than that of the unfrustrated one
[see the density of states (histogram, $\Delta E=0.02$) in the inset in the middle panel of Fig.~\ref{fig03}], 
thus leading to the drastic differences at low $T$, 
see the upper and middle panels of Fig.~\ref{fig03}.
Remarkably, the noticeable differences in all quantities are present at pretty high temperatures. 
Only, beyond $T \gtrsim 3$ the corresponding curves approach each other. 
A striking effect of frustration is also the shift of the maximum in $\chi(T)$ to lower temperatures, 
see the lower panel of Fig.~\ref{fig03}.
      
\subsection{Infinite lattice}
\label{sec4B}

Let us now move to a detailed investigation of the infinite PHAF by using the EM interpolation scheme, see Sec.~\ref{sec3A}.
Using our Magdeburg HTE code \cite{Lohmann2011,Lohmann-Diplom,Lohmann2014} 
we have created the HTE series for $c(T)$ and $\chi(T)$ up to order 13 (see Appendix~B)
that provides the high-temperature input for the EM. 
Since the EM finally uses Pad\'{e} approximants of a power series of $s(e)$ derived from the initial HTE series, 
the HTE input determines the highest order of the Pad\'{e} approximants of the EM interpolation scheme.    

As a low-temperature input for the EM we need the ground-state energy $e_0$.
There is a large variety of reported values for $e_0$ of the spin-half PHAF ranging from $e_0=-0.57$ to $e_0=-0.45$, 
namely,
$e_0=-0.572$ \cite{Sobral1997}, 
$-0.56$ \cite{Canals2000},
$-0.49$ \cite{Harris1991,Koga2001},
$-0.482081$ \cite{Chandra_ED_pyro_2018},
$-0.466971$ \cite{Chandra_ED_pyro_2018},
$-0.459$ \cite{Kim2008},
$-0.457804$ \cite{Isoda1998},    
$-0.45093$ \cite{RGM_pyro_2019},
$-0.4473$ \cite{Burnell2009}, 
i.e., accurate values for $e_0$ are missing.  
We also need the low-temperature law for $c(T)$, where we have to distinguish between gapped and gapless behavior
(cf. Sec.~\ref{sec3A}).
Finally, in case of gapless excitations the exponent $\alpha$ of the power law is an input parameter, 
and for the susceptibility $\chi_0\equiv\chi(T=0)$ is required as input.
Fortunately, for gapped excitations the value of the gap is not needed as an input, it is rather an output of the EM.
Moreover, we have $\chi_0\equiv\chi(T=0)=0$ in this case. 

We begin with the specific heat $c(T)$ for which the EM is better justified 
(two sum rules are exploited).
Actually, the low-temperature law for $c(T)$ is not known for the $S=1/2$ PHAF.
Advantageously, 
the previous ample of experience with the EM \cite{Bernu2001,Misguich2005,Bernu2013,Bernu2015,Schmidt_2017,Bernu2019}
provides valuable hints to overcome this difficulty.
Thus, 
in case that these input data are too far from the true (but possibly unknown) data one gets inconsistent or unphysical results.
Hence, 
to get physical (i.e., pole free) Pad\'{e} approximants requires reasonable values for $e_0$ and reasonable assumptions on the low-$T$ behavior of $c(T)$.
Moreover, 
getting a large number $n_{\rm P}$ of similar Pad\'{e} approximants for a certain input data set indicates the physical relevance of this set. 
On the other hand, 
the appearance of significant differences between the various Pad\'{e} approximants is a criterion to discard the corresponding input set.
The successfulness of this strategy has been demonstrated for several examples where excellent reference data are available,
in particular,
for the $S=1/2$ $XY$, HAF, and Ising chains \cite{Bernu2001,Bernu2015},
for the $S=1$ HAF (Haldane) chain \cite{Bernu2001},
for the $S=1/2$ square-lattice and triangular-lattice Heisenberg ferro- and antiferromagnets \cite{Bernu2001}
or
for the $S=1/2$ kagome-lattice HAFs \cite{Misguich2005,Bernu2015,Bernu2019}.

By combining different assumptions on the ground-state energy and low-$T$ behavior of $c(T)$ of the PHAF 
we have generated a large set of temperature profiles for the specific heat.
In the next step, we use the guidelines described above to evaluate the used input data, 
this way obtaining definite conclusions on their relevance.
In sum, a crucial criterion is that a particular input data set 
leads to a close bundle of temperature profiles obtained by many physical Pad\'{e} approximants.
Since, the high-temperature part is per se identical this criterion concerns the temperature region $T \lesssim 0.5$. 
In what follows, we will present in the main text only data for the most relevant input sets,
whereas presentation of some other illustrative results, 
only briefly mentioned in the main text, 
are transferred to Appendix~C.

\begin{figure}
\centering 
\includegraphics[clip=on,width=80mm,angle=0]{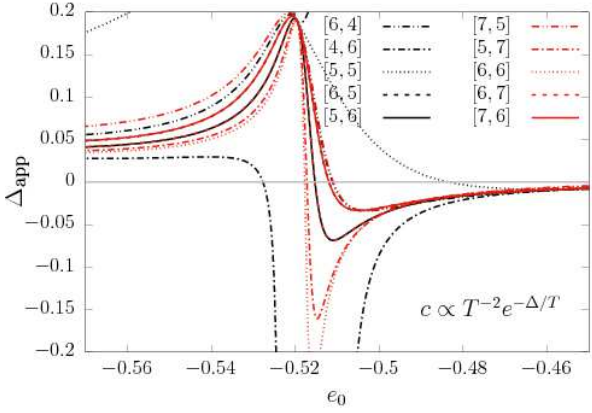}
\protect
\caption{EM for gapped spectrum: 
The gap $\Delta_{\rm app}$ as it follows from Eq.~(\ref{312}) 
versus 
the ground-state energy $e_0$ varied in region $-0.57\ldots-0.45$.}
\label{fig06} 
\end{figure}

\begin{figure}
\centering 
\includegraphics[clip=on,width=80mm,angle=0]{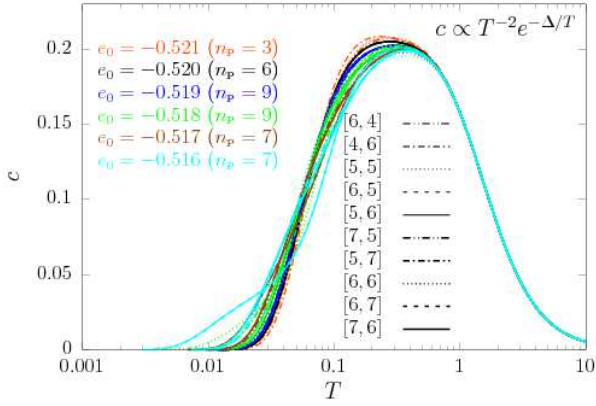}
\protect
\caption{EM results for the specific heat;
gapped spectrum, $e_0=-0.521\ldots-0.516$.
The parameter $n_{\rm P}$ in brackets behind energy values denotes the number of (very similar) Pad\'{e} approximants shown here.}
\label{fig07}  
\end{figure}

As a first result, 
we found that the assumption of gapped excitations is not favorable for the following reasons.
We varied $e_0$ in the region $-0.57\ldots-0.45$ and calculated the gap given in Eq.~(\ref{312}) using different Pad\'{e} approximants, 
where we focused on nearly diagonal Pad\'{e} approximants $[u, d](e)$, $u \sim d$,
constructed from HTE data of 10th, 11th, 12th, and 13th order,
see Fig.~\ref{fig06}.
We find that the gap $\Delta_{\rm app}$, Eq.~(\ref{312}), is negative if the ground-state energy exceeds approximately $-0.515$,
thus providing evidence that a gapped spectrum together with $e_0 \gtrsim -0.515$ can be excluded.
For ground-state energies in the region $-0.519 \ldots -0.517$ we obtain $\Delta_{\rm app} = 0.16 \ldots 0.18$,
see Fig.~\ref{fig06} (and Fig.~\ref{fig16} in Appendix~C),
and there is a decent number $n_{\rm P}$ of Pad\'{e} approximants yielding similar $c(T)$ profiles,
see Fig.~\ref{fig07}.
For $e_0$ less than $-0.521$ the number of physical Pad\'{e} approximants noticeably decreases. 
Although the results for $c(T$) do not entirely discard a gapped ground state, 
further EM analysis of $\chi(T)$ under this assumption leads to disagreement with diagrammatic Monte Carlo simulations of Ref.~\cite{Huang2016} at $T<0.7$,
see Fig.~\ref{fig13} below.
We may consider these findings for $c(T$) and $\chi(T)$ as an indication to favor a gapless ground state.
We mention that the Green's function results indicate gapless magnetic excitations \cite{RGM_pyro_2019}, 
and, as mentioned already above,  
the data for $\chi(T)$ given in \cite{Huang2016} down to $T=J/6$ seem also to be in favor of gapless excitations.

\begin{figure}
\centering 
\includegraphics[clip=on,width=80mm,angle=0]{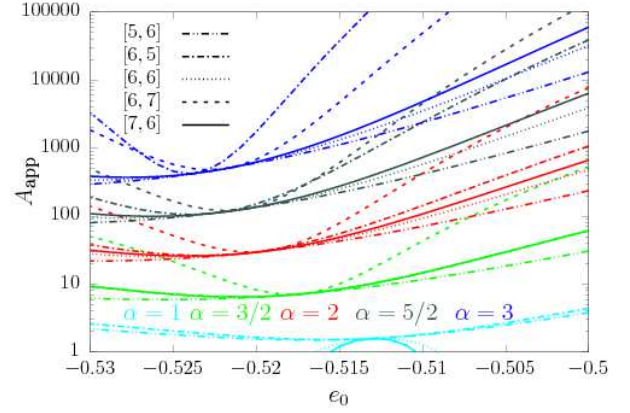}
\protect
\caption{EM for gapless spectrum: 
The prefactor $A_{\rm app}$ as it follows from Eq.~(\ref{308}) for $\alpha=1,3/2,2,5/2,3$ 
versus 
the ground-state energy $e_0$ varied in the region $-0.53\ldots-0.50$.}
\label{fig08}
\end{figure}

We focus now on the gapless case with a power-law decay of the specific heat $c(T)=AT^{\alpha}$ as $T\to 0$.
Since we do not know the exponent $\alpha$, we study different values $\alpha=1, 3/2, 2, 5/2, 3$.
As for the gapped case,
we again varied $e_0$ in the region $-0.57\ldots-0.45$.
We observed that only in a much smaller region around $e_0=-0.52$ reasonable results can be obtained 
(see also the discussion of Fig.~\ref{fig08}, below). 
Thus, in what follows we consider preferably the region $-0.53\ldots-0.50$ in more detail.

First we consider the prefactor $A_{\rm app}$ that is given within the EM by Eq.~(\ref{308}).
According to above outlined criteria for physically relevant EM outcomes
the values of $A_{\rm app}$ obtained by different Pad\'{e} approximants must be very close to each other.   
From Fig.~\ref{fig08} (see also Fig.~\ref{fig17} in Appendix~C for additional information) it is evident
that for each value of $\alpha$ there is a well-defined relevant region of $e_0$, 
namely,
$-0.515 \ldots -0.513$ for $\alpha=1$,
$-0.518 \ldots -0.516$ for $\alpha=3/2$,
$-0.521 \ldots -0.518$ for $\alpha=2$,
$-0.522 \ldots -0.520$ for $\alpha=5/2$,
$-0.524 \ldots -0.522$ for $\alpha=3$.
In all cases the ground-state energy is within the interval $e_0=-0.524\ldots-0.513$,
which is much narrower than the wide region reported in the literature ranging from $e_0=-0.57$ to $e_0=-0.45$
\cite{Sobral1997,Canals2000,Harris1991,Koga2001,Chandra_ED_pyro_2018,Kim2008,Isoda1998,RGM_pyro_2019,Burnell2009}.

\begin{figure}
\centering 
\includegraphics[clip=on,width=80mm,angle=0]{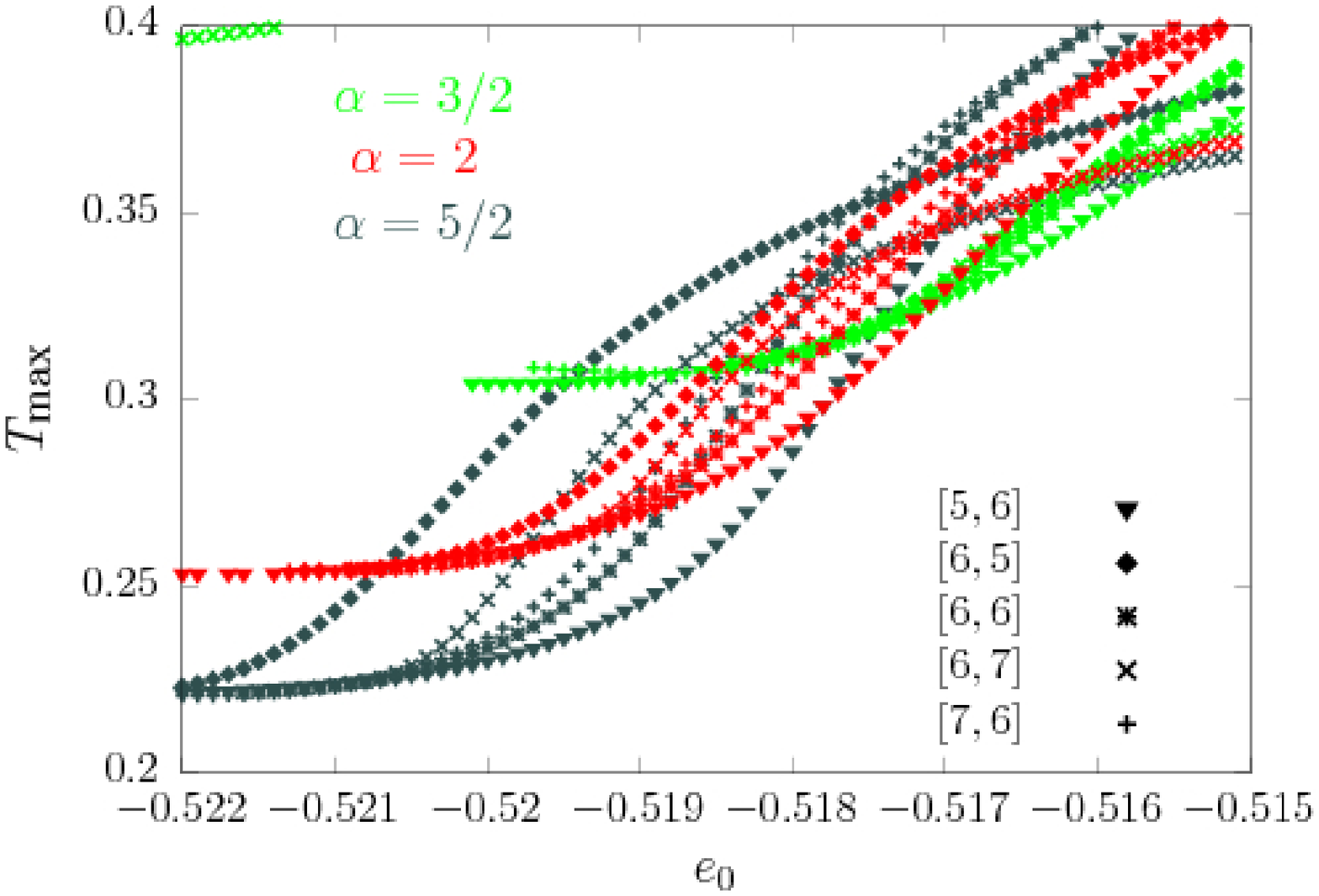}
\includegraphics[clip=on,width=80mm,angle=0]{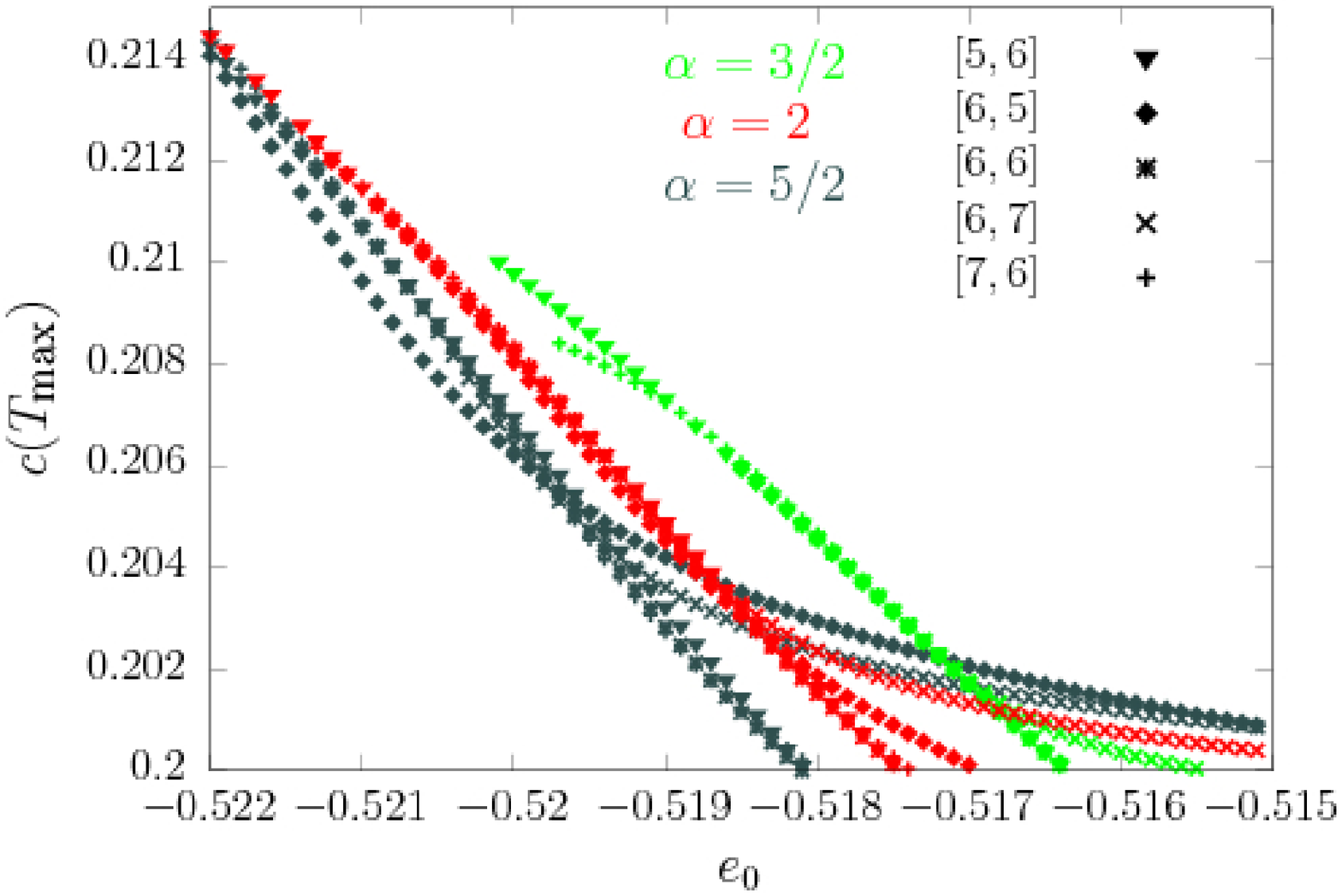}
\protect
\caption{EM for gapless spectrum:
Position $T_{\max}$ (top) and height $c(T_{\max})$ (bottom) of the specific heat 
versus 
the ground-state energy $e_0$ varied in the region region $-0.522\ldots-0.515$ 
for exponents $\alpha=3/2$ (green), $\alpha=2$ (red), and $\alpha=5/2$ (dark gray).}
\label{fig09}
\end{figure}

We consider the coincidence of $A$ values for different Pad\'{e} approximants as a necessary criterion 
to figure out regions of relevant values for $e_0$ and $\alpha$. 
Since $A$ (together with $\alpha$) determines the $c(T)$ profile at sufficiently low $T$, 
we can get additional reliability by examining the region around the main (``high''-temperature) maximum. 
For that 
we compare the position $T_{\max}$ and the height $c(T_{\max})$ of this maximum obtained from different Pad\'{e} approximants 
in dependence on $e_0$ within the regions guided by the previous inspection of $A$ for $\alpha=3/2,2,5/2$ in Fig.~\ref{fig09} 
(Fig.~\ref{fig18} in Appendix~C reports such data for a wider region of $e_0$ including also $\alpha=1$ and $3$).
For each $\alpha$ we find pretty small regions of $e_0$ which yield almost identical $T_{\max}$ and $c(T_{\max})$, 
and, consistently, this region fits well to the region obtained by inspection of $A$.
For example, 
for $\alpha=2$ all Pad\'{e} approximants give almost identical $T_{\max}$ and $c(T_{\max})$, cf. Fig.~\ref{fig09} (red symbols),
if $e_0$ is taken within the region $-0.522\ldots-0.519$, 
which is in excellent agreement with that obtained from the prefactor $A$, see above.  
Note, however, that for $\alpha=1$ and $3$ the diversity of $T_{\max}$ and $c(T_{\max})$ is noticeably larger than for $\alpha=3/2,2,5/2$, 
cf. Fig.~\ref{fig18} in Appendix~C, 
indicating that the exponents $\alpha=1$ and $3$ are less favorable.

\begin{figure}
\centering 
\includegraphics[clip=on,width=80mm,angle=0]{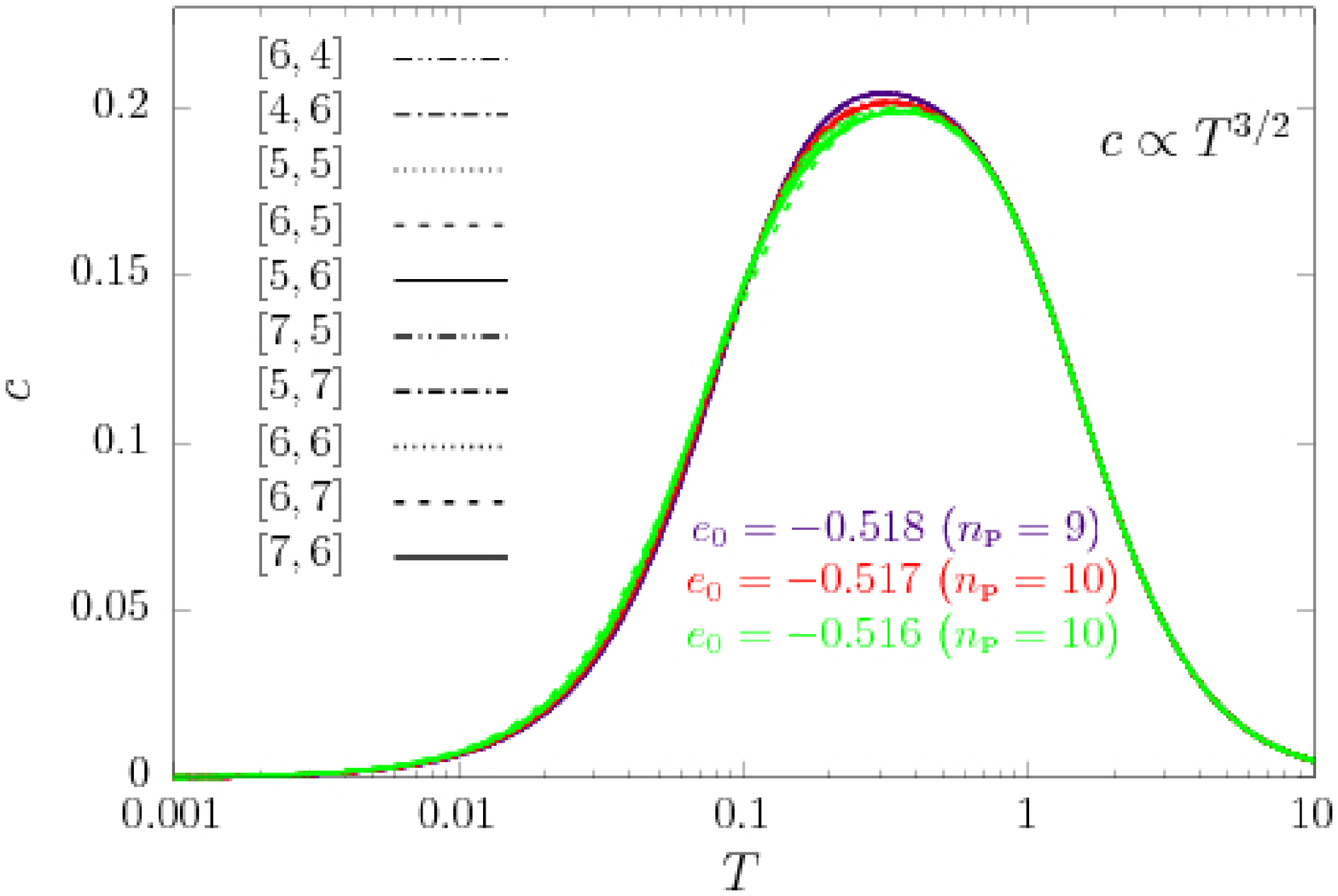} 
\includegraphics[clip=on,width=80mm,angle=0]{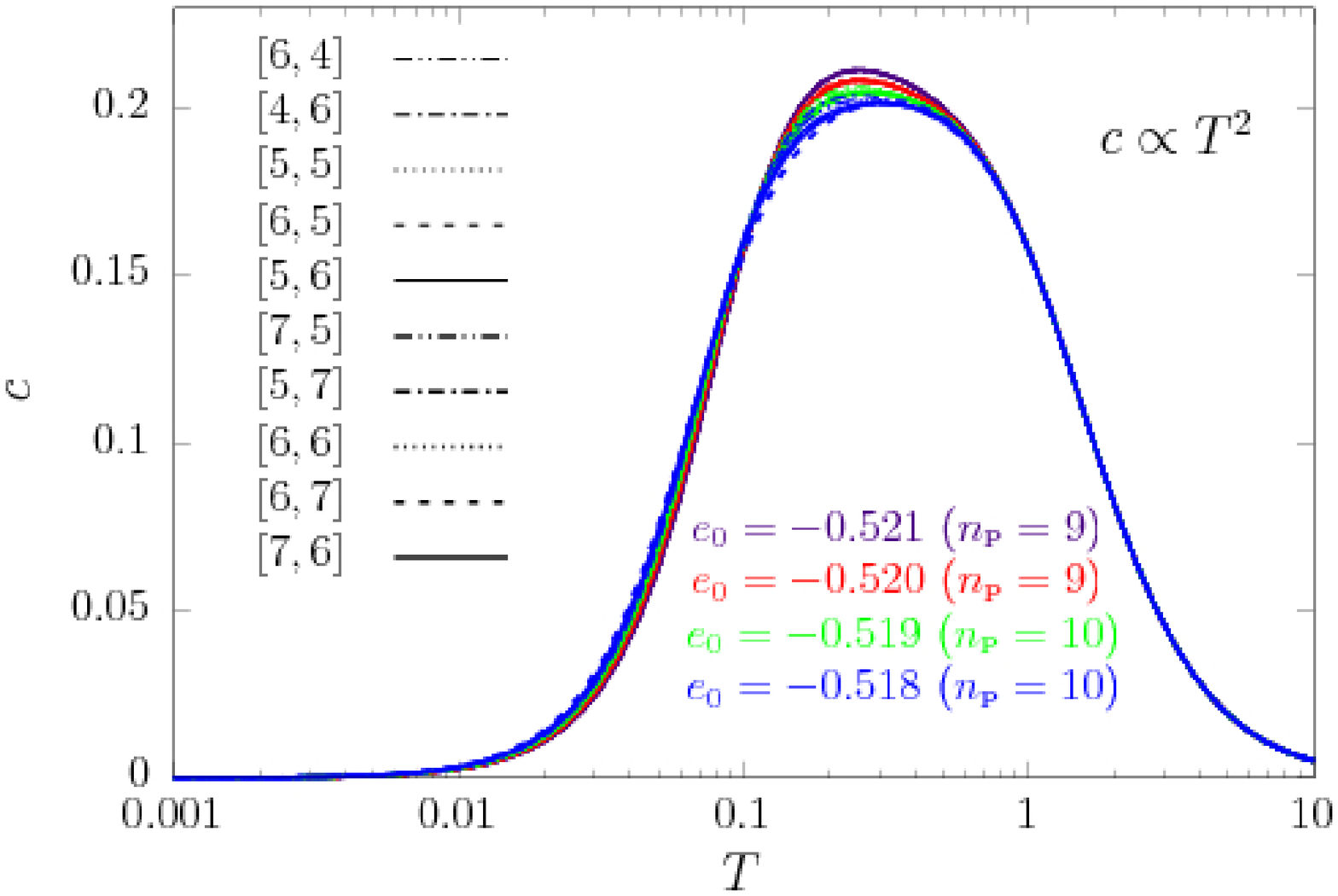} 
\includegraphics[clip=on,width=80mm,angle=0]{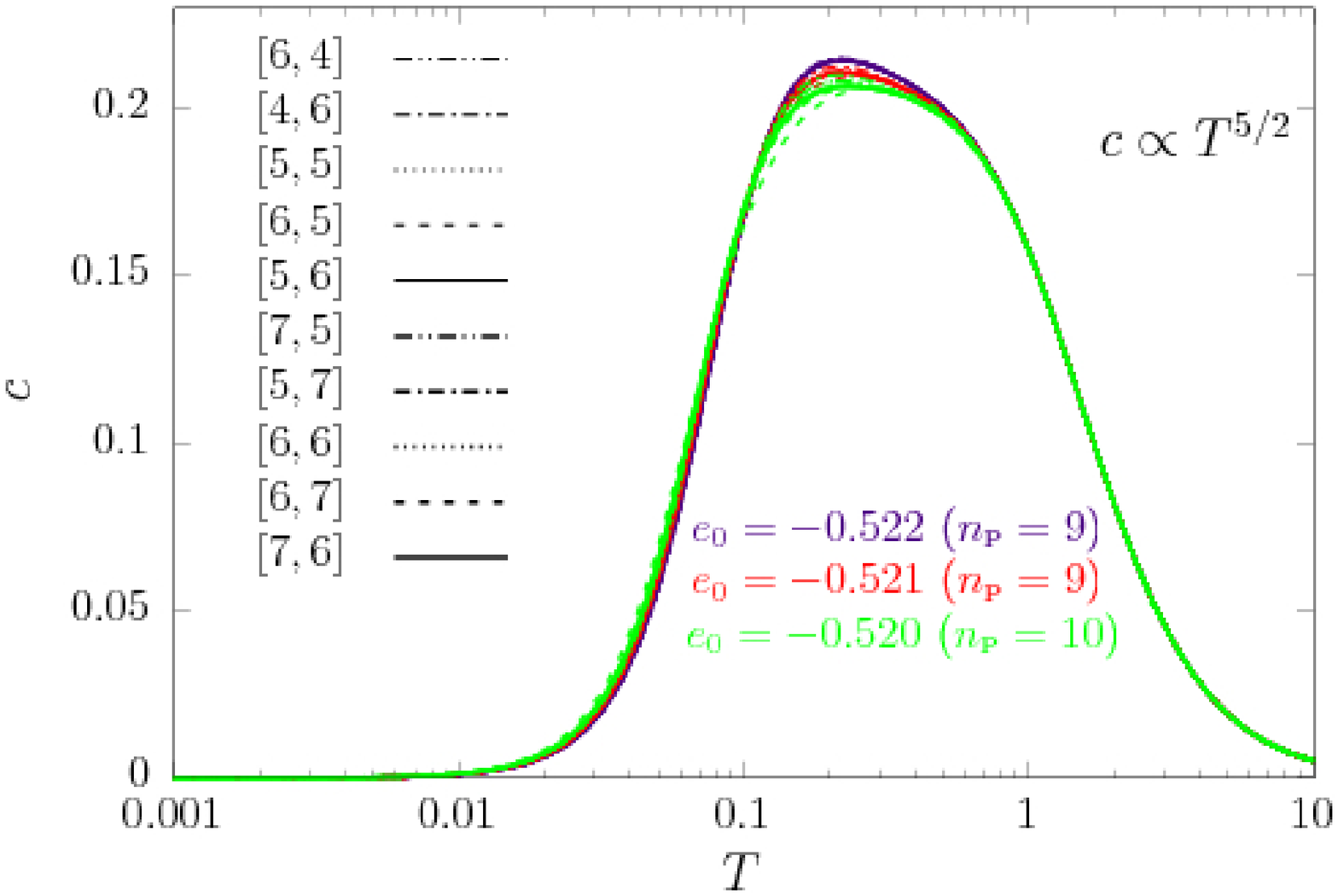} 
\protect
\caption{EM results for the specific heat;
gapless spectrum.
(Top) $e_0=-0.518\ldots -0.516$, $\alpha=3/2$.
(Middle) $e_0=-0.521\ldots -0.518$, $\alpha=2$.
(Bottom) $e_0=-0.522\ldots -0.520$, $\alpha=5/2$.
The parameter $n_{\rm P}$ in brackets behind energy values denotes the number of (very similar) Pad\'{e} approximants shown here.} 
\label{fig10} 
\end{figure}

Finally, 
after the specification of the ground-state energy values as outlined above,
we present in Fig.~\ref{fig10} the full $c(T)$ curves obtained by the EM for $\alpha=3/2,2,5/2$ and a few related optimal values of $e_0$.  
[Corresponding curves for $\alpha=1$ and $\alpha=3$ are shown in Fig.~\ref{fig19} in Appendix~C.
Moreover, 
Fig.~\ref{fig20} in Appendix~C provides additional results of $c(T)$ comparing data for all $\alpha=1,3/2,2,5/2,3$ for various values of $e_0$.]
 
As can be seen in Fig.~\ref{fig10}, 
there is a quite large number of Pad\'{e} approximants 
(see the parameter $n_{\rm P}$ in brackets behind energy values)
yielding very similar temperature profiles $c(T)$. 
Thus, for $\alpha=2$ we show in the middle panel of Fig.~\ref{fig10}
$n_{\rm P}=10$ Pad\'{e} approximants if $e_0=-0.519$ and $-0.518$ 
and
$n_{\rm P}=9$ Pad\'{e} approximants if $e_0=-0.521$ and $e_0=-0.520$.
Outside this region of $e_0$ values the number of physical Pad\'{e} approximants becomes noticeably smaller.

Without favoring any of the assumptions on the low-$T$ behavior of $c(T)$
and taking into account all (i.e., gapped and gapless excitations) EM predictions for $c(T)$
collected in Figs.~\ref{fig07} and \ref{fig10}
(see also Fig.~\ref{fig21} in Appendix~C, where we present a direct comparison of both cases),
we have clear evidence 
1) for a quite narrow region of reasonable $e_0$ values 
and
2) for the absence of a double-peak profile in $c(T)$.
3) Though, the position $T_{\max}$ and the height $c(T_{\max})$ of the maximum of the specific heat 
slightly depend on the assumption about the ground-state energy $e_0$ and low-lying excitations, 
all cases yield $T_{\max}$ around $\approx 0.25$ and $c(T_{\max})$ around $\approx 0.2$.
  
\begin{figure}
\centering 
\includegraphics[clip=on,width=80mm,angle=0]{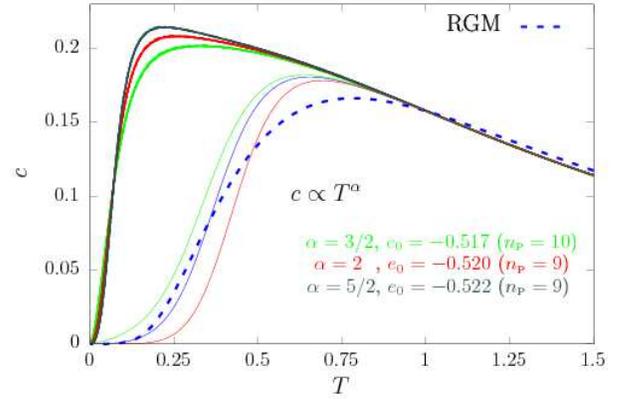}
\protect
\caption{Specific heat $c(T)$ of the $S=1/2$ PHAF:
Comparison of our EM data ($e_0=-0.522$, $\alpha=5/2$; $e_0=-0.520$, $\alpha=2$; $e_0=-0.517$, $\alpha=3/2$) 
with results obtained 
by the rotation-invariant Green's function method \cite{RGM_pyro_2019} (dashed blue)
and 
by the HTE (thin solid curves; we show the same HTE data for $N=\infty$ as shown in Fig.~\ref{fig04}).}
\label{fig11}
\end{figure}

Finally, we compare our EM results for $c(T$) 
with data calculated by HTE (without subsequent EM interpolation) and by the Green's function approach \cite{RGM_pyro_2019}, 
cf. Fig.~\ref{fig11}.
The Green's function results deviate from the EM results already below $T \sim 1$, 
whereas the HTE data deviate below $T \sim 0.7$.

\begin{figure}
\centering 
\includegraphics[clip=on,width=80mm,angle=0]{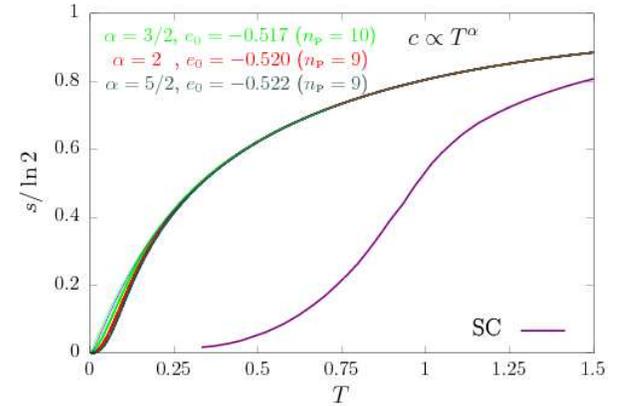}
\protect
\caption{EM data for the entropy for the gapless case with $e_0=-0.522$, $\alpha=5/2$; $e_0=-0.520$, $\alpha=2$; $e_0=-0.517$, $\alpha=3/2$.
We also show quantum Monte-Carlo data for the simple-cubic HAF \cite{sc_entropy_2010} as well as HTE data (thin solid lines),
where we show the same Pad\'{e} approximants as used in Fig.~\ref{fig04} for $c(T)$.  
(Note the the HTE data are very close to the EM data.)} 
\label{fig12}
\end{figure}

The EM straightforwardly also provides the temperature profile of the entropy $s(T)$, see Fig.~\ref{fig12}. 
Since the finite-temperature phase transition present in the simple-cubic HAF does not influence $s(T)$ as much as $c(T)$,
we compare the data for the PHAF with corresponding ones for the simple-cubic HAF taken from Ref.~\cite{sc_entropy_2010}. 
We also show HTE data.
Similar as already found for the finite system, cf. the middle panel of Fig.~\ref{fig03}, 
the frustration leads to a much faster increase of $s$ at low temperatures for the PHAF. 
Thus, at $T=0.5$ the entropy already amounts to more than 50\% of its maximal value $\ln2 \approx 0.69$.
(Note that in Fig.~\ref{fig22} in Appendix~C we present a direct comparison of the gapped and gapless temperature profiles of $s$.)

\begin{figure}
\centering 
\includegraphics[clip=on,width=80mm,angle=0]{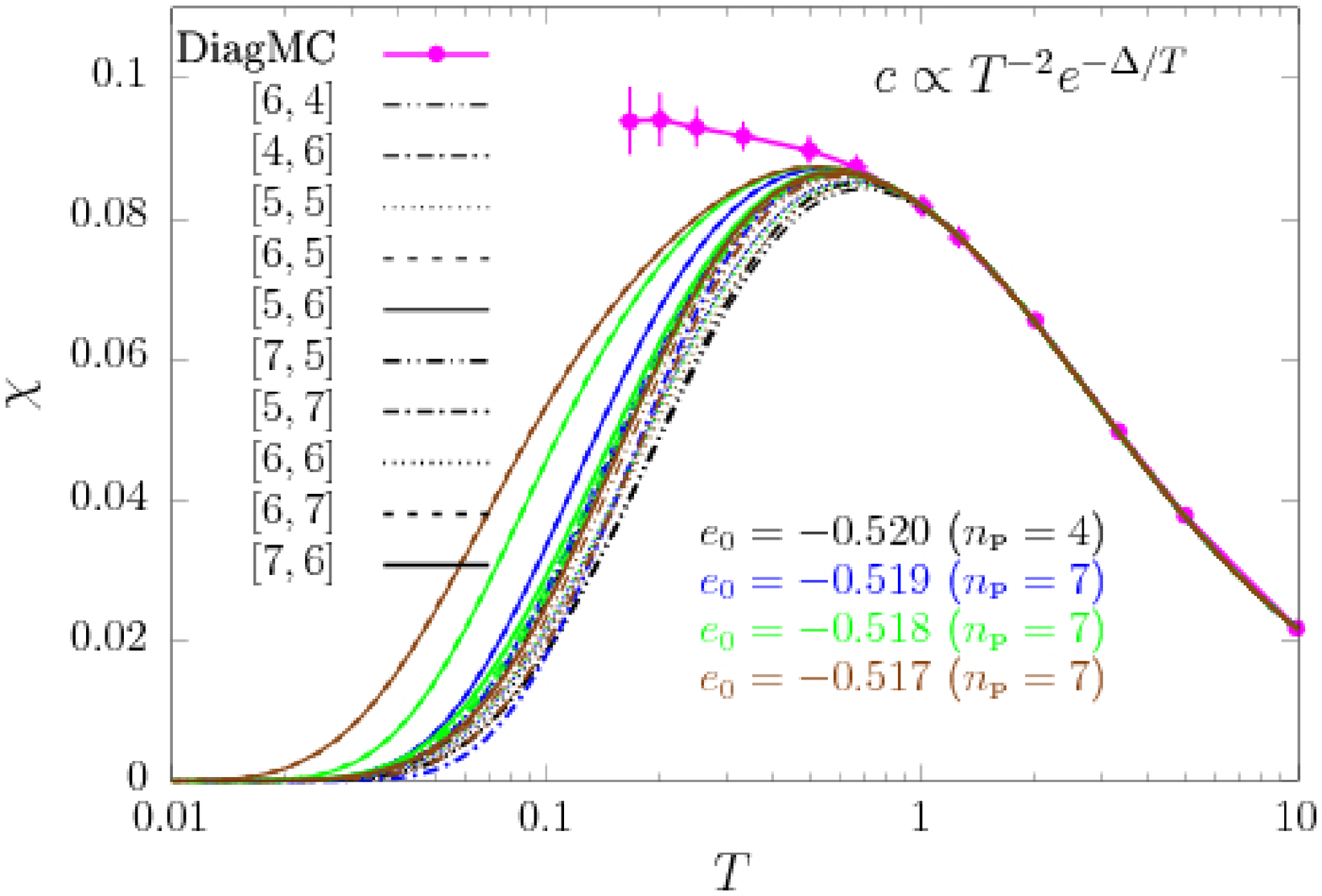}
\protect
\caption{EM results for the susceptibility; gapped spectrum, $e_0=-0.520\ldots-0.517$.}
\label{fig13}
\end{figure}

\begin{figure}
\centering 
\includegraphics[clip=on,width=80mm,angle=0]{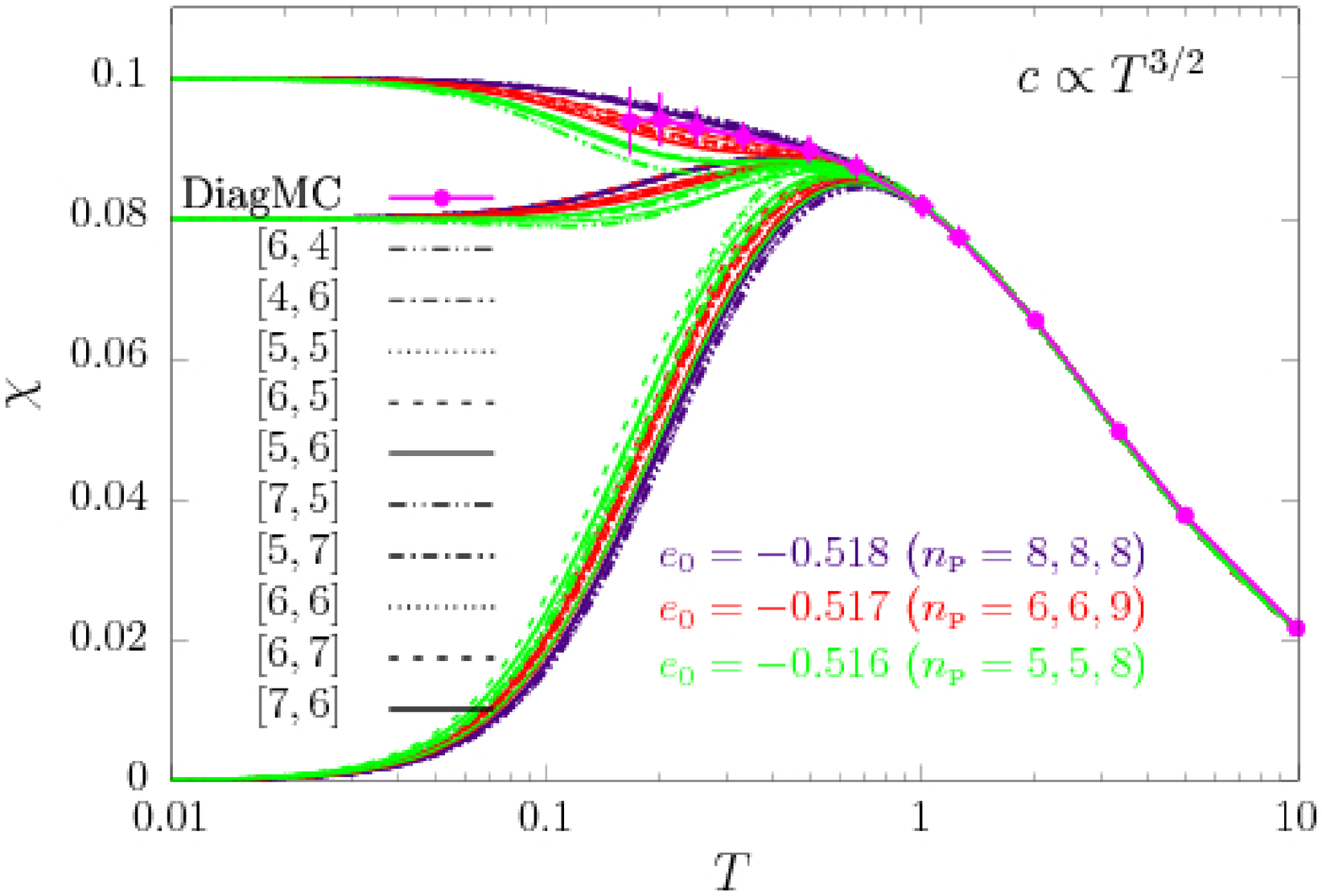} 
\includegraphics[clip=on,width=80mm,angle=0]{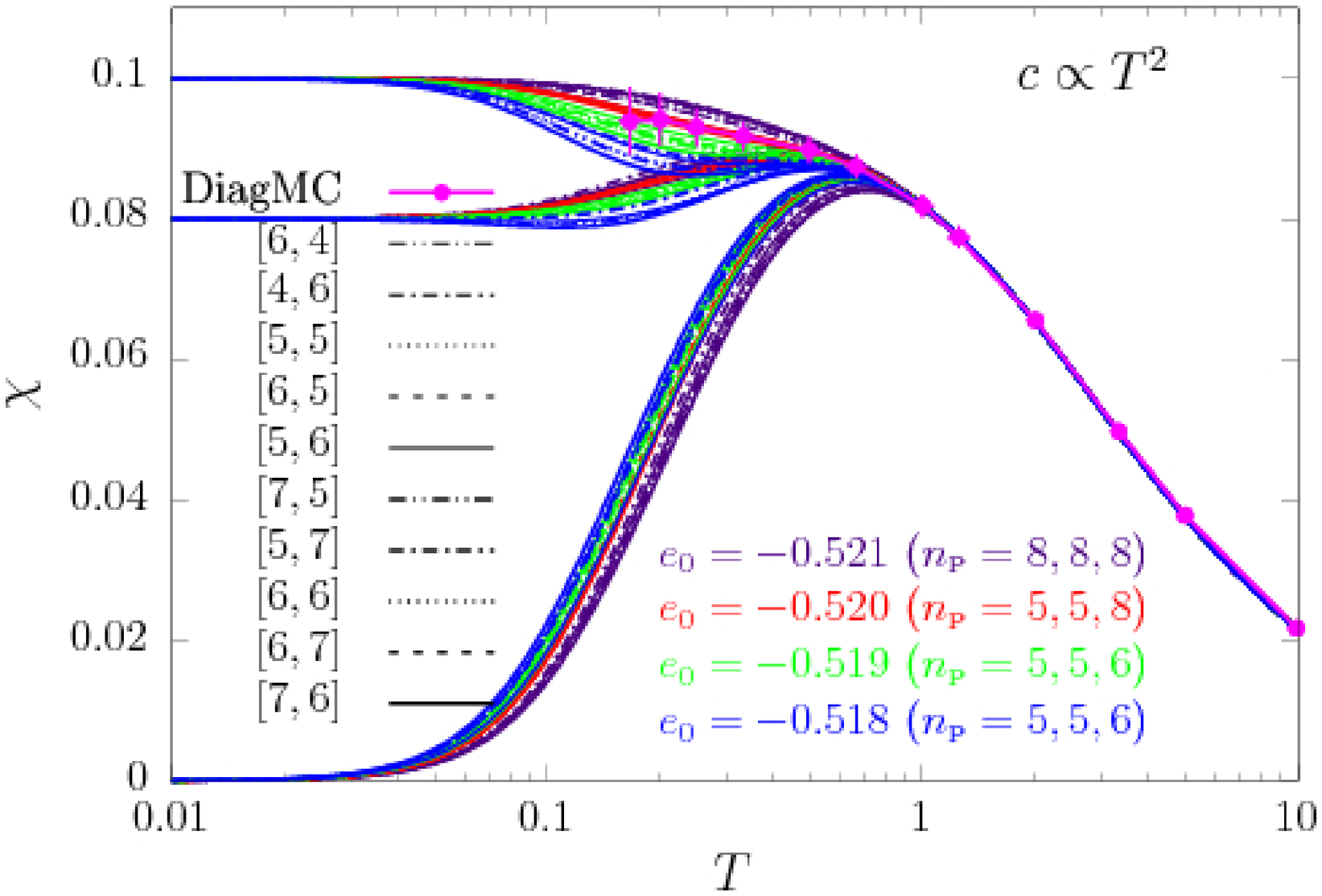} 
\includegraphics[clip=on,width=80mm,angle=0]{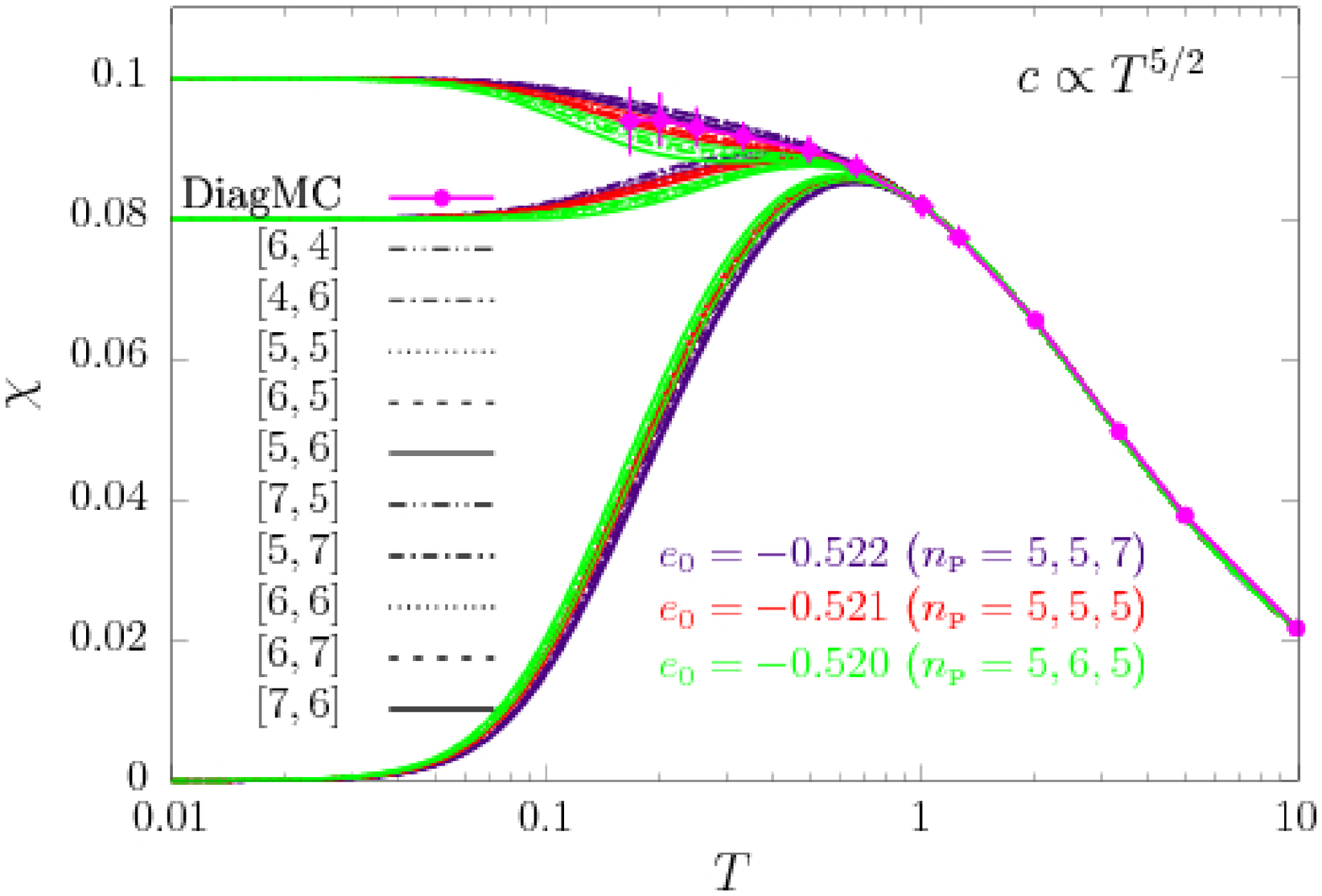} 
\protect
\caption{EM results for the susceptibility; gapless spectrum.
(Top) $e_0=-0.518\ldots -0.516$,  $\alpha=3/2$.
(Middle) $e_0=-0.521\ldots -0.518$,  $\alpha=2$.
(Bottom) $e_0=-0.522\ldots -0.520$,  $\alpha=5/2$.
$\chi_0=0,0.08,0.1$.
Three values of $n_{\rm P}$ in brackets correspond to the assumed values $\chi_0=0.1$, $\chi_0=0.08$, and $\chi_0=0$, consequently.}
\label{fig14} 
\end{figure}

We consider now the static uniform susceptibility  $\chi$ calculated by the EM as described in Sec.~\ref{sec3A}.
According to the results of the rotation-invariant Green's function method, there is $\chi_0\equiv\chi(T=0)>0$. 
A non-zero $\chi_0$ may be also expected from the diagrammatic Monte Carlo simulations \cite{Huang2016}.
Nevertheless, we will not exclude from the beginning $\chi_0=0$, i.e., a non-zero spin gap.
Although the above discussed EM data for $c(T)$ are in favor of a gapless spectrum,  
the specific heat profiles were not fully conclusive to entirely reject the gapped spectrum.
Moreover, one could have gapless singlet (i.e., non-magnetic) excitations but gapped triplet (i.e., magnetic) excitations.
Thus we studied the case with exponential decay of $\chi(T)$ and $c(T)$ as $T\to 0$ 
(i.e., gapped singlet and triplet excitations), 
see Fig.~\ref{fig13},
as well as the  case with exponential decay of $\chi(T)$ and power-law decay of $c(T)$ as $T\to 0$,
(i.e., gapless singlet and gapped triplet excitations), 
see Fig.~\ref{fig14}, 
where we consider those values for $e_0$ and $\alpha$ which are previously used to get $c(T)$
(see also Figs.~\ref{fig23} and \ref{fig24} in Appendix~C).  
From these figures, 
it is obvious that the susceptibility profiles based on gapped magnetic excitations 
are not compatible with the diagrammatic Monte Carlo data of Ref.~\cite{Huang2016} in the temperature region below $0.7$, 
thus providing further evidence against a gapped spectrum.
On the other hand,
the EM results for gapless excitations with $\alpha=3/2,2,5/2$ and $e_0=-0.522\ldots-0.516$ and with non-zero $\chi_0$ 
as shown in Fig.~\ref{fig14}
fit much better to the data of Ref.~\cite{Huang2016}.

\begin{figure}
\centering 
\includegraphics[clip=on,width=80mm,angle=0]{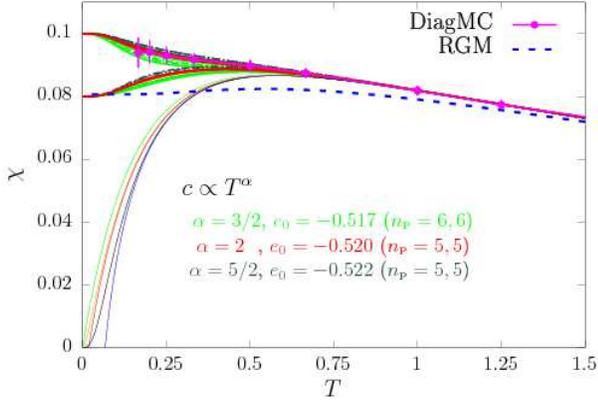}
\protect
\caption {Susceptibility $\chi(T)$ of the $S=1/2$ PHAF:
Comparison of our EM data 
($e_0=-0.522$, $\alpha=5/2$; $e_0=-0.520$, $\alpha=2$; $e_0=-0.517$, $\alpha=3/2$) 
with results obtained 
by the rotation-invariant Green's function method \cite{RGM_pyro_2019} (dashed blue),    
by diagrammatic Monte Carlo \cite{Huang2016},
and 
by the HTE (thin solid curves; we show the same HTE data for $N=\infty$ as shown in Fig.~\ref{fig05}).}
\label{fig15}
\end{figure}

As for the specific heat, 
we finally compare our EM results for $\chi(T$) with data calculated 
by HTE (without subsequent EM interpolation), 
by the Green's function approach \cite{RGM_pyro_2019}, 
cf. Fig.~\ref{fig15},
where we also include the data of the diagrammatic Monte Carlo simulations \cite{Huang2016}.
The Green's function results deviate already below $T \sim 1.5$, and the HTE coincides down to  $T \sim 0.7$, 
whereas $\chi(T)$ profiles with $\chi_0=0.1$ are in excellent agreement with the data of Ref.~\cite{Huang2016}.

\section{Conclusions and summary}
\label{sec5}
\setcounter{equation}{0}

We have studied the specific heat $c(T)$, the entropy $s(T)$, and the static uniform susceptibility $\chi(T)$ 
of the spin-half pyrochlore Heisenberg antiferromagnet (PHAF)
on a finite lattice of $N=32$ sites using the finite-temperature Lanczos method (FTLM)  
and 
on the infinite lattice using the high-temperature expansion (HTE) up to order 13 
and a sophisticated interpolation 
between the low- and high-temperature behavior of the thermodynamic potential entropy $s$ as a function of internal energy $e$ 
[the  entropy method (EM)].

We found that finite lattices of such size are not appropriate to get reasonable results below $T \sim 0.7$, 
but they might be useful 
to get a general impression on the strong frustration effects present in the PHAF by comparison of the HAF on finite pyrochlore and simple-cubic lattices.
A similar limitation to temperatures above $T \sim 0.7$ is valid for the HTE
even if the range of validity of the high-temperature series is extended by Pad\'{e} approximants.
Only the EM interpolation is suitable to overcome these limitation and to provide sound data for the whole temperature range. 

Our main findings for the specific heat $c(T)$ are as follows.
(i) Contrary to the two-dimensional kagome HAF, 
we do not find hints neither for an extra low-$T$ peak nor an extra shoulder below the main maximum.
However, 
the absence of an extra low-$T$ feature goes hand in hand with a significant shift of the single maximum towards $T \sim 0.25$,
which is much lower than for the kagome HAF,
where the main maximum is at $T_{\rm max}/J = 0.67$ \cite{kago42,Bernu2019}.    
This conclusion is robust,
i.e., it is obtained not only for gapless excitations for all reasonable exponents $\alpha$ of the low-temperature power law of $c(T)$, 
but holds also for gapped excitations.
(ii) A gapless spectrum is more favorable than a gapped one, 
i.e., most likely there is power-law low-$T$ behavior of $c(T)$. 
Although best results are for an exponent $\alpha=2$, 
other exponents ($\alpha=1,3/2,5/2,3$) cannot be excluded. 
(iii) We predict a ground-state energy $e_0\approx -0.52$ \footnote{It is worth to make the following remark here.
As explained in the main text, we followed the protocol to determine the best ground-state energy suggested in Ref.~\cite{Bernu2019}
see Sec.~IIIE in Supplemental Material of arXiv:1909.00993v1.
The authors of that paper tested this protocol and commented on its accuracy 
that is important, in particular, when the HTE is known at not very high orders.}.

Our EM data for the susceptibility $\chi(T)$ in comparison with data obtained by diagrammatic Monte Carlo \cite{Huang2016} 
provide further evidence for a gapless spectrum with a ground-state energy $e_0\approx-0.52$.
The temperature profile of $\chi$ most likely does not show a maximum, 
rather there is a monotonous increase of $\chi$ upon decreasing of $T$ reaching a zero-temperature value of $\chi_0 \approx 0.1$.

Recently, we have learned that R.~Sch\"{a}fer et al. \cite{Schaefer2020} also examined thermodynamics of the quantum PHAF, 
however, using for that numerical linked-cluster expansions.

\section*{Acknowledgments}

The authors gratefully acknowledge helpful discussions with David~J.~Luitz and Imre~Hagym\'{a}si.
They thank Paul~A.~McClarty for his critical reading of the manuscript and helpful comments and suggestions.
O.~D. acknowledges the kind hospitality of the MPIPKS, Dresden in September-November of 2019.
T.~H. was supported 
by the State Fund for Fundamental Research of Ukraine (project F82/45950 ``Effects of frustration in quantum spin systems'') 
and 
by the fellowship of the National Academy of Sciences of Ukraine for young scientists.

\section*{Appendix A: M.~Planck's derivation of the specific heat $c(T)$ of an oscillator}
\renewcommand{\theequation}{A.\arabic{equation}}
\setcounter{equation}{0}

In his revolutionary paper in 1900 \cite{Planck1}
(see  also the Nobel Prize address 
[Notes 7, 12, and 13 at the end of the Nobel Prize address 
``The origin and development of the quantum theory'' by Max Planck
delivered before the Royal Swedish Academy of Sciences at Stockholm, 2 June, 1920] \cite{Planck2}),
M.~Planck investigated the energy distribution of electromagnetic radiation emitted by a black body in thermal equilibrium. 
For that he considered the entropy of the equilibrium radiation $S$ \footnote{In Eqs.~(\ref{a01}) -- (\ref{a05}) we use notations of Ref.~\cite{Planck2}.} 
in relation with its energy $U$,
or more accurately,
the second derivative ${\rm d}^2S/{\rm d} U^2$. 
According to Wien's law it is 
\begin{eqnarray}
\label{a01}
\frac{{\rm d}^2S}{{\rm d} U^2}=-\frac{1}{bU}.
\end{eqnarray}
But in view of experiments for high temperatures one has $U=cT$, 
i.e.,
\begin{eqnarray}
\label{a02}
\frac{{\rm d}^2S}{{\rm d} U^2}=-\frac{c}{U^2}. 
\end{eqnarray}
(M.~Planck here refers to experiments by F.~Kurlbaum.)
While Wien's law (\ref{a01}) is valid for small energy values (short wave length),
Eq.~(\ref{a02}) describes the high-energy limit (long wave length, Rayleigh-Jeans law).  
To get agreement with experimental data M.~Planck suggested 
\begin{eqnarray}
\label{a03}
\frac{{\rm d}^2S}{{\rm d} U^2}
=-\frac{c}{U\left(bc+U\right)}
\longrightarrow
\left\{
\begin{array}{cc}
-\frac{1}{bU},  & U \ll bc, \\
-\frac{c}{U^2}, & U \gg bc,
\end{array}
\right.
\end{eqnarray}
which interpolates between both limiting cases.
By integrating we get
\begin{eqnarray}
\label{a04}
\frac{1}{T}=\frac{{\rm d}S}{{\rm d} U}=\frac{1}{b}\ln\left(1+\frac{bc}{U}\right),
\end{eqnarray}
which yields Planck's formula
\begin{eqnarray}
\label{a05}
U=\frac{bc}{e^{b/T}-1}.
\end{eqnarray}

These arguments can be formulated within the setup of the entropy method 
to find the specific heat $c(T)$ of a (Bose) oscillator,
which represents the electromagnetic radiation with the frequency $\nu=2\pi\omega$.
Now we know that $b=\hbar\omega$ and $c=1$.
Taking into account the zero-point energy we have to replace $U\to e=U+\hbar\omega/2$. 
Then the Planck's interpolation formula (\ref{a03}) for the auxiliary function $G(e)=(s(e))^{\prime\prime}$ reads:
\begin{eqnarray}
\label{a06}
G(e)=-\frac{1}{\left(e-\frac{\hbar\omega}{2}\right)\left(e+\frac{\hbar\omega}{2}\right)}
=\frac{1}{\left(\frac{\hbar\omega}{2}\right)^2-e^2}
\nonumber\\
\longrightarrow
\left\{
\begin{array}{cc}
-\frac{1}{\hbar\omega \left(e-\frac{\hbar\omega}{2}\right)},  & e \ll \hbar\omega, \\
-\frac{1}{e^2}, & e \gg \hbar\omega.
\end{array}
\right.
\end{eqnarray}
The subscript {\tiny{app}} in the left-hand side of this equation is omitted 
since the suggested expression for $G(e)=(s(e))^{\prime\prime}$ (\ref{a06}) appears to be exact.
By integrating we get
\begin{eqnarray}
\label{a07}
\frac{1}{T} = s^{\prime}(e) 
=\int\limits_e^\infty{\rm d}e\frac{1}{e^2-\left(\frac{\hbar\omega}{2}\right)^2}
=\frac{1}{\frac{\hbar\omega}{2}}{\rm arccth}\frac{e}{\frac{\hbar\omega}{2}}
\end{eqnarray}
and then
\begin{eqnarray}
\label{a08}
c(e)=-\frac{(s^\prime(e))^2}{s^{\prime\prime}(e)}
=
\frac{e^2-\left(\frac{\hbar\omega}{2}\right)^2}{\left(\frac{\hbar\omega}{2}\right)^2}{\rm arccth}^2\frac{e}{\frac{\hbar\omega}{2}}
\end{eqnarray}
and finally
\begin{eqnarray}
\label{a09}
c(T)=\left(\frac{\frac{\hbar\omega}{2T}}{{\rm sh}\frac{\hbar\omega}{2T}}\right)^2.
\end{eqnarray}

\section*{Appendix B: HTE for the $S=1/2$ PHAF}
\renewcommand{\theequation}{B.\arabic{equation}}
\setcounter{equation}{0}

We report here the HTE coefficients for the spin-half PHAF up to order 13 
obtained by the Magdeburg HTE code developed mainly by A.~Lohmann \cite{Lohmann-Diplom}.
Note that the coefficients up to order 10 where presented previously for arbitrary spin $S$ in Refs.~\cite{Lohmann2011,Lohmann2014}.
    
For the specific heat (per site) we have
\begin{eqnarray}
\label{b01}
c(\beta)=\sum_{i}d_i\beta^i,  
\nonumber\\
d_{1}=0,
\;
\frac{d_{2}}{J^2}=+\frac{9}{16}, 
\;
\frac{d_{3}}{J^3}=-\frac{9}{32}, 
\;
\frac{d_{4}}{J^4}=-\frac{207}{256}, 
\nonumber\\
\frac{d_{5}}{J^5}=+\frac{195}{256}, 
\;
\frac{d_{6}}{J^6}=+\frac{3\,549}{4\,096}, 
\nonumber\\
\frac{d_{7}}{J^7}=-\frac{59\,073}{40\,960}, 
\;
\frac{d_{8}}{J^8}=-\frac{34\,535}{65\,536}, 
\nonumber\\
\frac{d_{9}}{J^9}=+\frac{345\,491}{163\,840}, 
\;
\frac{d_{10}}{J^{10}}=-\frac{9\,385\,203}{36\,700\,160}, 
\nonumber\\
\frac{d_{11}}{J^{11}}=-\frac{337\,285\,883}{132\,120\,576}, 
\;
\frac{d_{12}}{J^{12}}=+\frac{39\,036\,781\,051}{26\,424\,115\,200}, 
\nonumber\\
\frac{d_{13}}{J^{13}}=+\frac{144\,963\,365\,443}{58\,133\,053\,440}. \quad 
\end{eqnarray}

For the static uniform susceptibility (per site) we have
\begin{eqnarray}
\label{b02}
\chi(\beta)=\sum_{i}c_i\beta^i,  
\nonumber\\
c_{1}=+\frac{1}{4}, 
\;
\frac{c_{2}}{J}=-\frac{3}{8}, 
\;
\frac{c_{3}}{J^2}=+\frac{3}{8}, 
\;
\frac{c_{4}}{J^3}=-\frac{17}{64}, 
\nonumber\\
\frac{c_{5}}{J^4}=+\frac{85}{512}, 
\;
\frac{c_{6}}{J^5}=-\frac{97}{640}, 
\nonumber\\
\frac{c_{7}}{J^6}=+\frac{20\,207}{122\,880}, 
\;
\frac{c_{8}}{J^7}=-\frac{210\,989}{1\,720\,320}, 
\nonumber\\
\frac{c_{9}}{J^8}=+\frac{92\,147}{1\,966\,080}, 
\;
\frac{c_{10}}{J^9}=-\frac{4\,936\,709}{247\,726\,080}, 
\nonumber\\
\frac{c_{11}}{J^{10}}=+\frac{540\,939\,383}{9\,909\,043\,200}, 
\;
\frac{c_{12}}{J^{11}}=-\frac{5\,315\,724\,257}{72\,666\,316\,800}, 
\nonumber\\
\frac{c_{13}}{J^{12}}=+\frac{20\,479\,483\,351}{747\,424\,972\,800}.   \quad
\end{eqnarray}

\section*{Appendix C: Additional EM data for $c$, $s$, and $\chi$ of the $S=1/2$ PHAF}
\renewcommand{\theequation}{C.\arabic{equation}}
\setcounter{equation}{0}

In this appendix we collect more figures presenting EM data for the specific heat, the entropy, and the susceptibility 
which are briefly discussed but not shown as figures in Sec.~\ref{sec4B}. 

\begin{figure}
\centering 
\includegraphics[clip=on,width=80mm,angle=0]{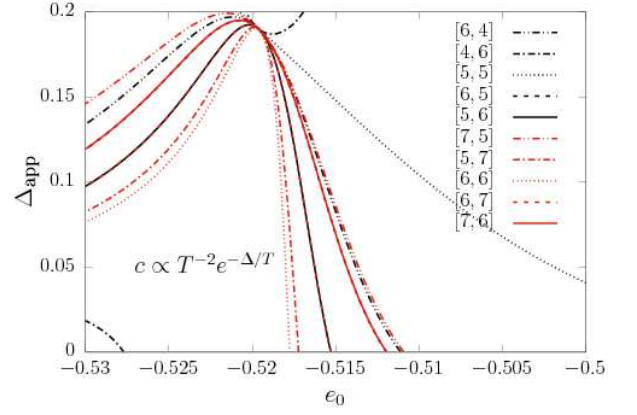}
\protect
\caption{The same as in Fig.~\ref{fig06}, but for $e_0$ values within a narrower region $-0.53\ldots-0.50$.}
\label{fig16}
\end{figure}

In Fig.~\ref{fig16} we show some results related to Fig.~\ref{fig06} using a smaller region of $e_0$.

\begin{figure}
\centering 
\includegraphics[clip=on,width=80mm,angle=0]{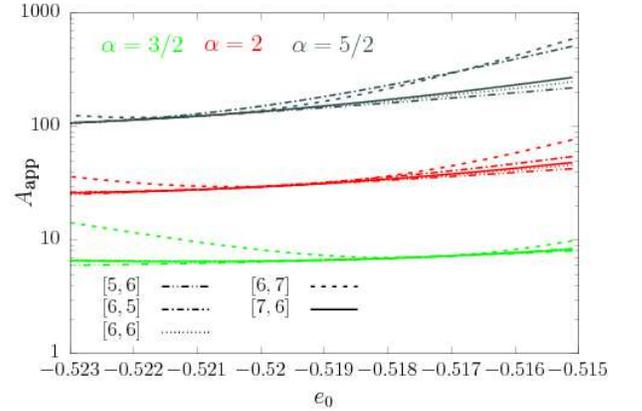}
\protect
\caption{The same as in Fig.~\ref{fig08}, but for $\alpha=3/2,2,5/2$ only and a narrower region of $e_0=-0.523\ldots-0.515$.}
\label{fig17} 
\end{figure}

In Fig.~\ref{fig17} we show some results related to Fig.~\ref{fig08} using a smaller region of $e_0$.
The region of $e_0$ in which various Pad\'{e} approximants yield almost the same prefactor $A_{\rm app}$ (\ref{308}) is different for $\alpha=3/2,2,5/2$.
Clearly, the values of $\alpha$ and $A$ are linked.
For example,
after assuming $\alpha=2$ for $e_0=-0.520\ldots-0.519$
the EM prediction for the specific heat as $T\to 0$ reads:
$c(T)=A_{\rm app}T^2$ with $A_{\rm app}=29 \ldots 31$.

\begin{figure}
\centering 
\includegraphics[clip=on,width=80mm,angle=0]{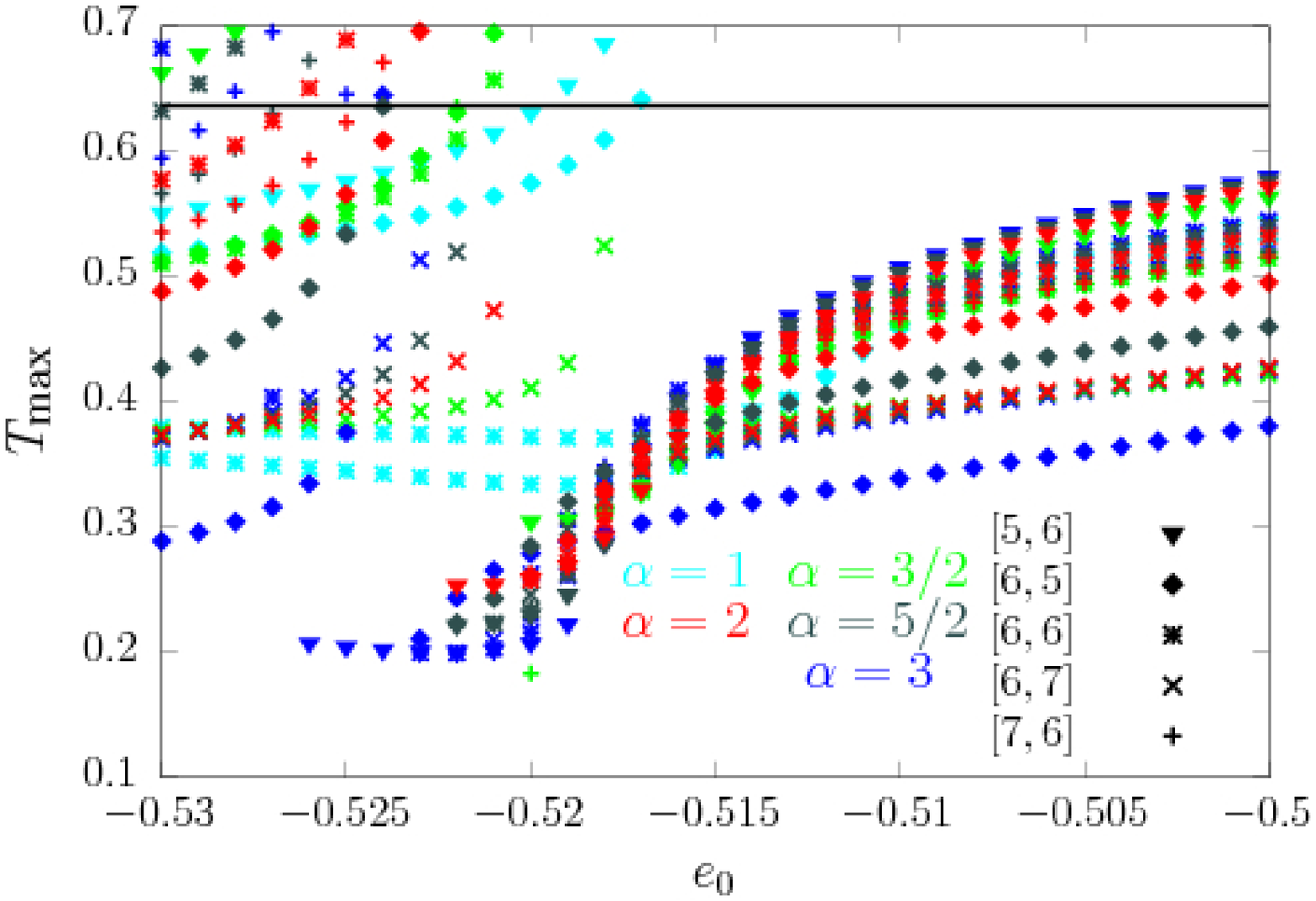}
\includegraphics[clip=on,width=80mm,angle=0]{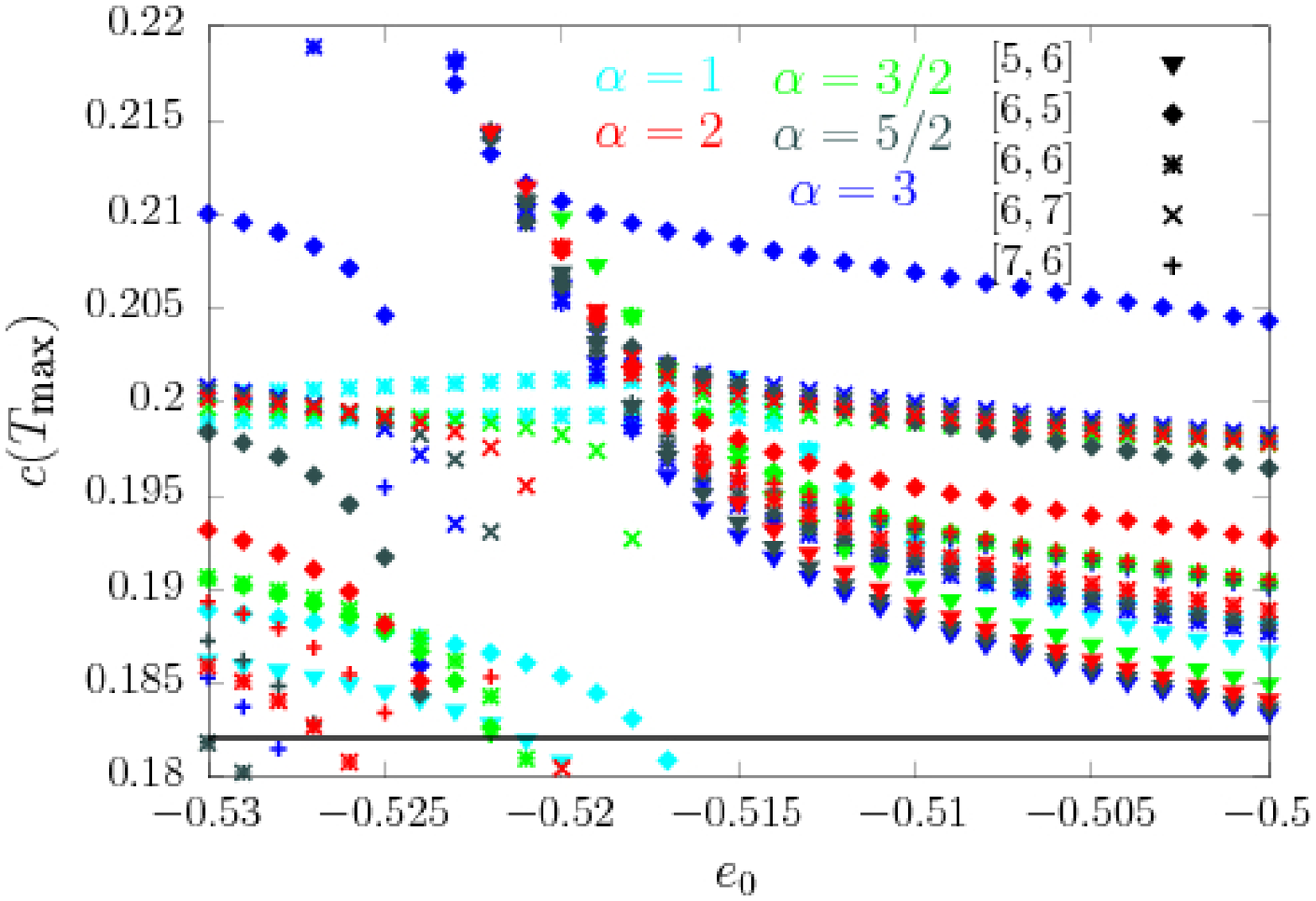} 
\protect
\caption{The same as in Fig.~\ref{fig09}, but for wider region of $e_0=-0.53\ldots-0.5$ and all five values of $\alpha=1,3/2,2,5/2,3$.
Black straight lines correspond to the Pad\'{e} approximant $[6,7]$ of high-temperature series of $c(T)$
[$T_{\max}\approx 0.637$, $c(T_{\max})\approx 0.182$].}
\label{fig18} 
\end{figure}

In Fig.~\ref{fig18} we show some results related to Fig.~\ref{fig09} using a wider region of $e_0$ and including exponents $\alpha=1$ and $\alpha=3$.
The position and the height of the maximum of the specific heat, as they follow from raw HTE series extended by the Pad\'{e} approximant $[6,7]$,
have the values $T_{\max}\approx 0.637$ and $c(T_{\max})\approx 0.182$.
The EM predictions for $\alpha=2$ and $e_0=-0.520\ldots-0.519$ are different:
$T_{\max}\approx 0.25\ldots0.28$ and $c(T_{\max})\approx 0.205\ldots0.208$.

\begin{figure}
\centering 
\includegraphics[clip=on,width=80mm,angle=0]{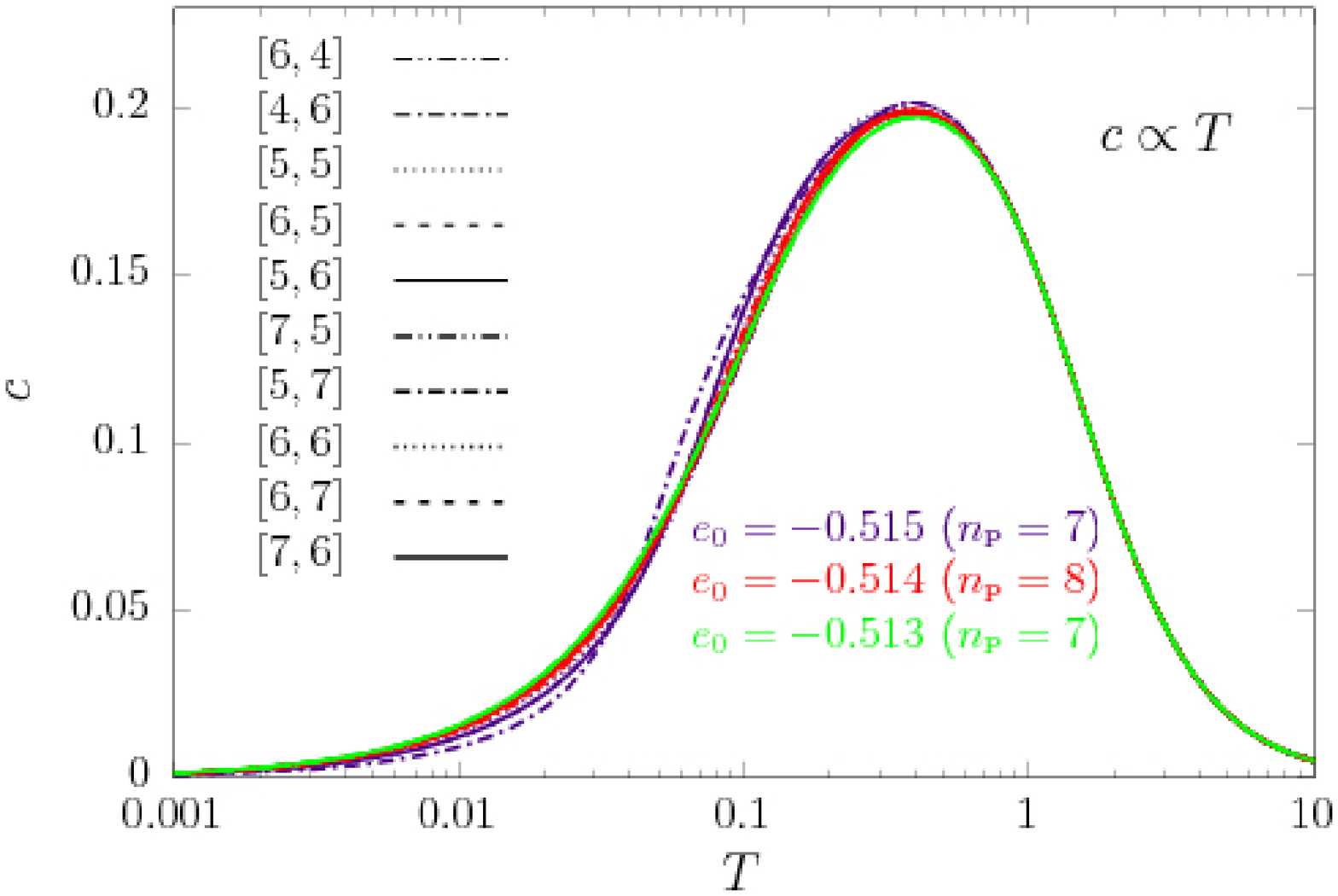}
\includegraphics[clip=on,width=80mm,angle=0]{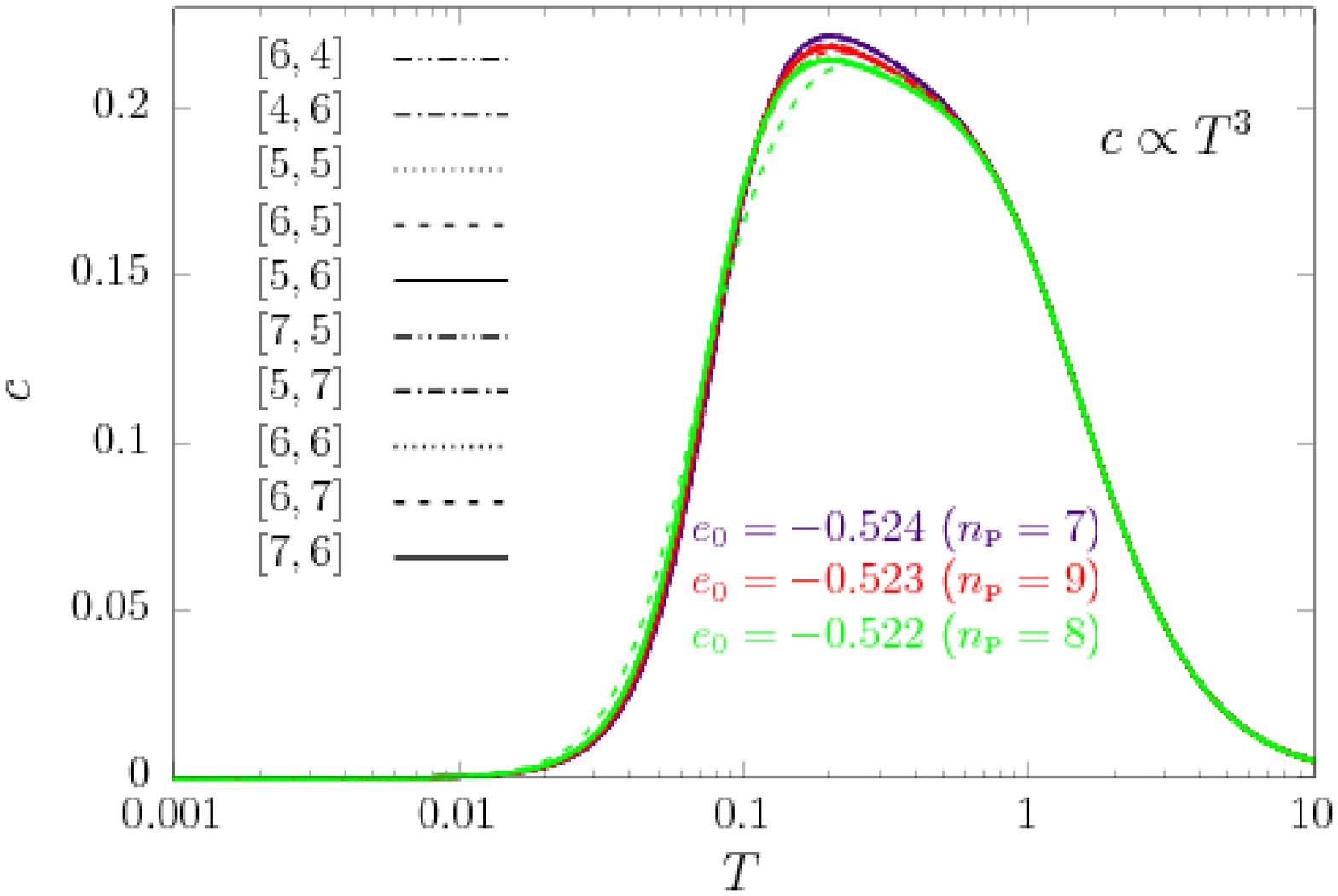}
\protect
\caption{Supplement to Fig.~\ref{fig10}; 
EM results for the specific heat;
gapless spectrum.
(Top) $e_0=-0.515\ldots -0.513$,  $\alpha=1$.
(Bottom) $e_0=-0.524\ldots -0.522$,  $\alpha=3$.}
\label{fig19}
\end{figure}

Fig.~\ref{fig19} is supplementary to Fig.~\ref{fig10} of the main text.
It contains similar EM predictions for $c(T)$ under less favorable assumptions $\alpha=1$ and $\alpha=3$.

\begin{figure}
\centering 
\includegraphics[clip=on,width=80mm,angle=0]{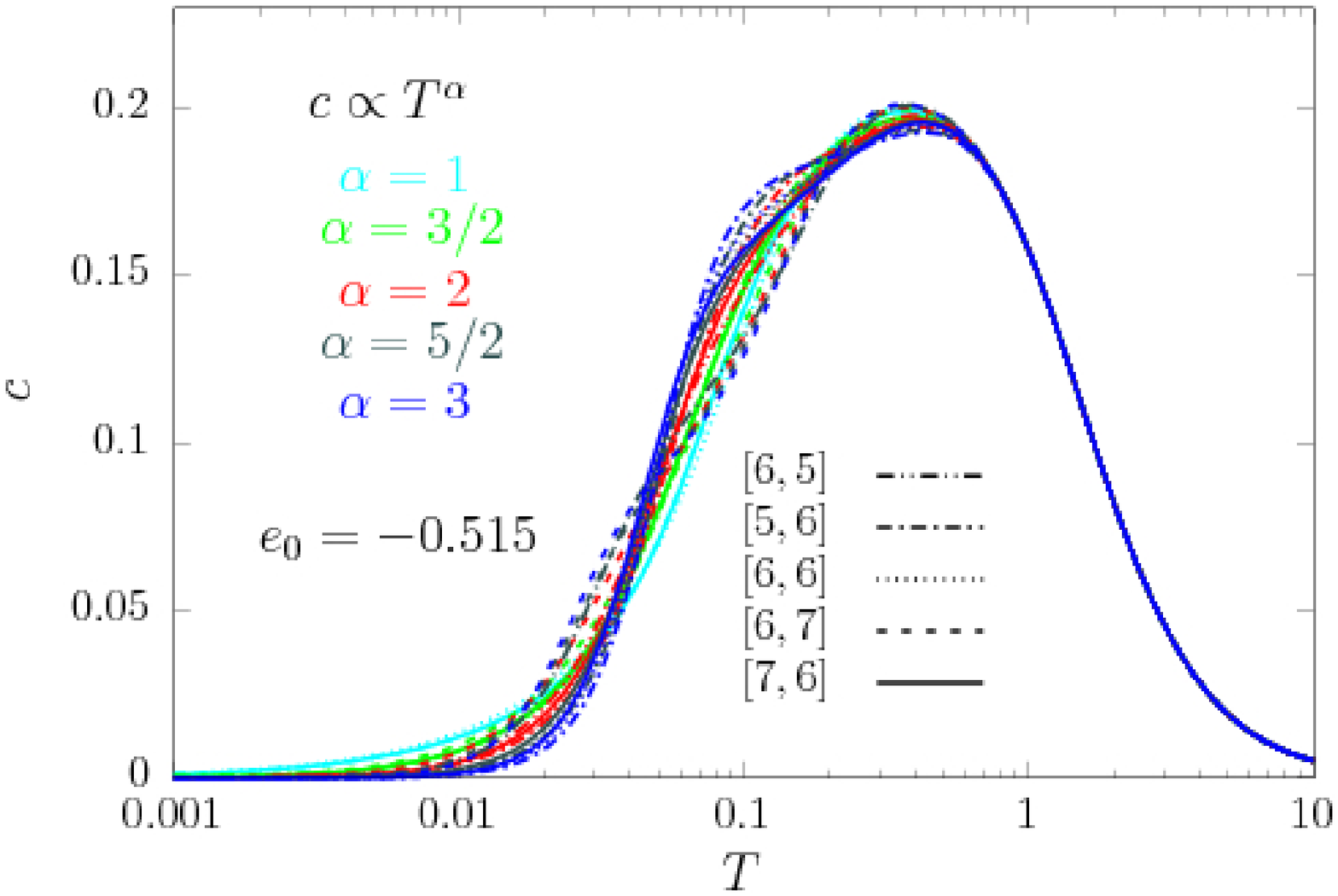} 
\includegraphics[clip=on,width=80mm,angle=0]{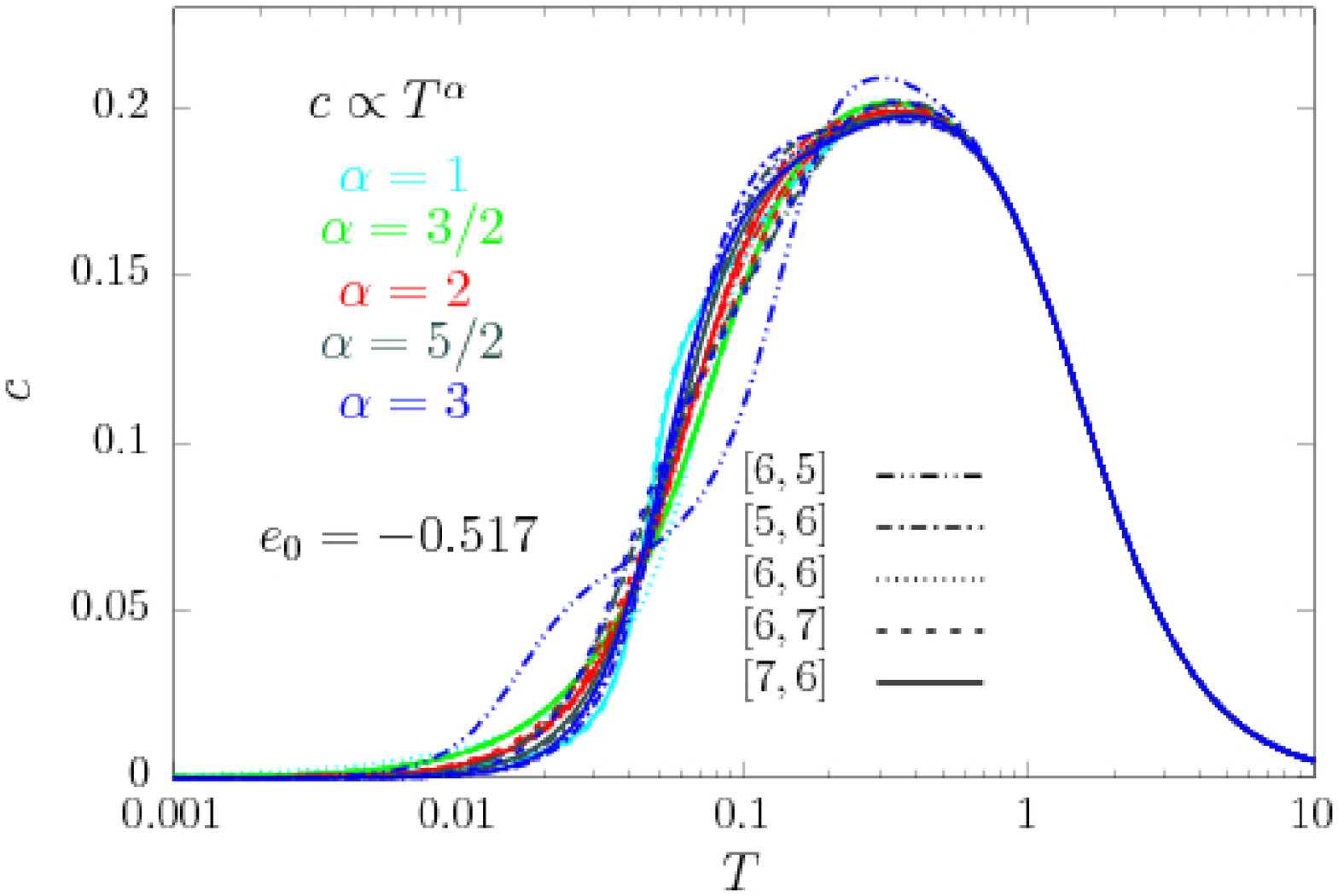} 
\includegraphics[clip=on,width=80mm,angle=0]{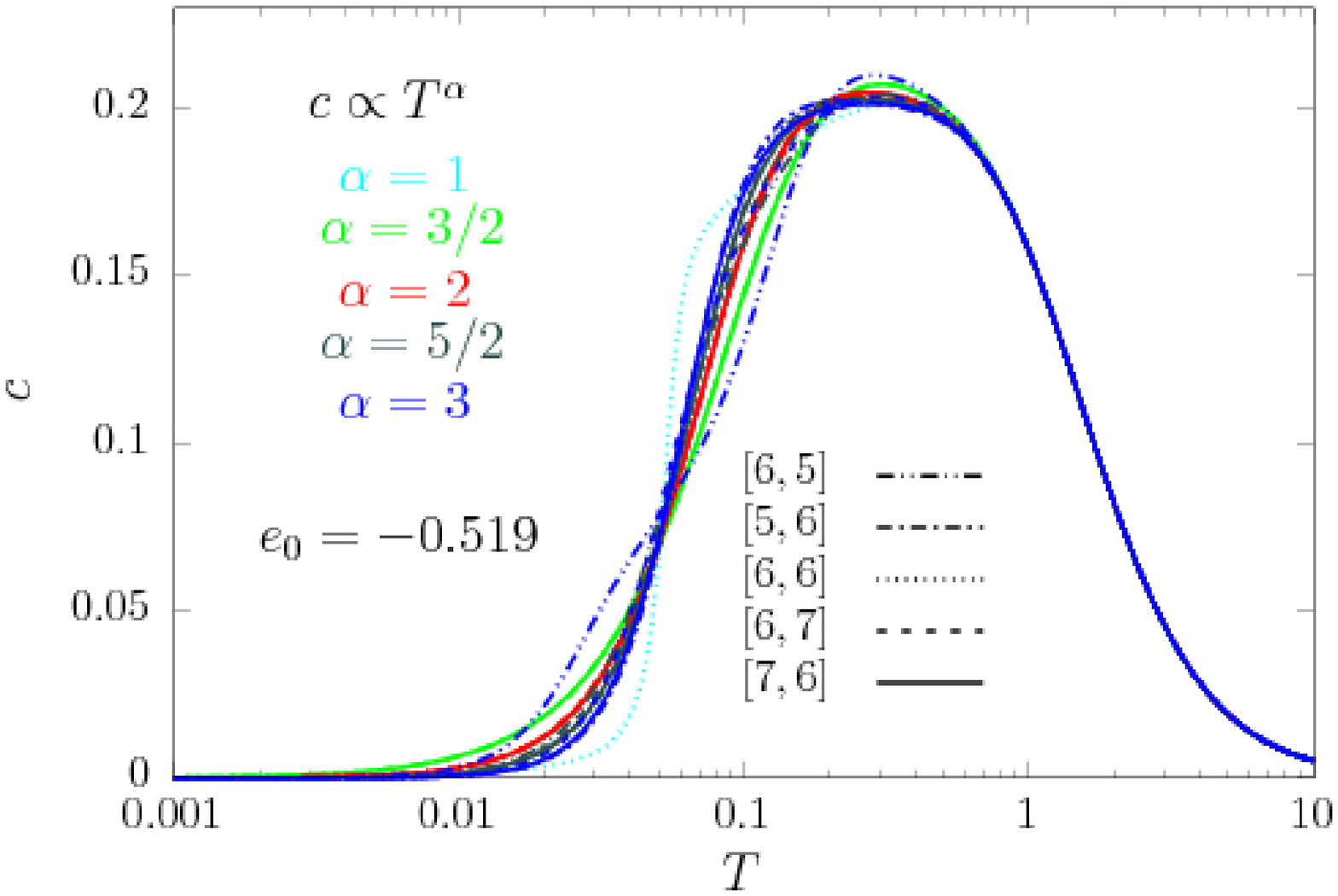} 
\includegraphics[clip=on,width=80mm,angle=0]{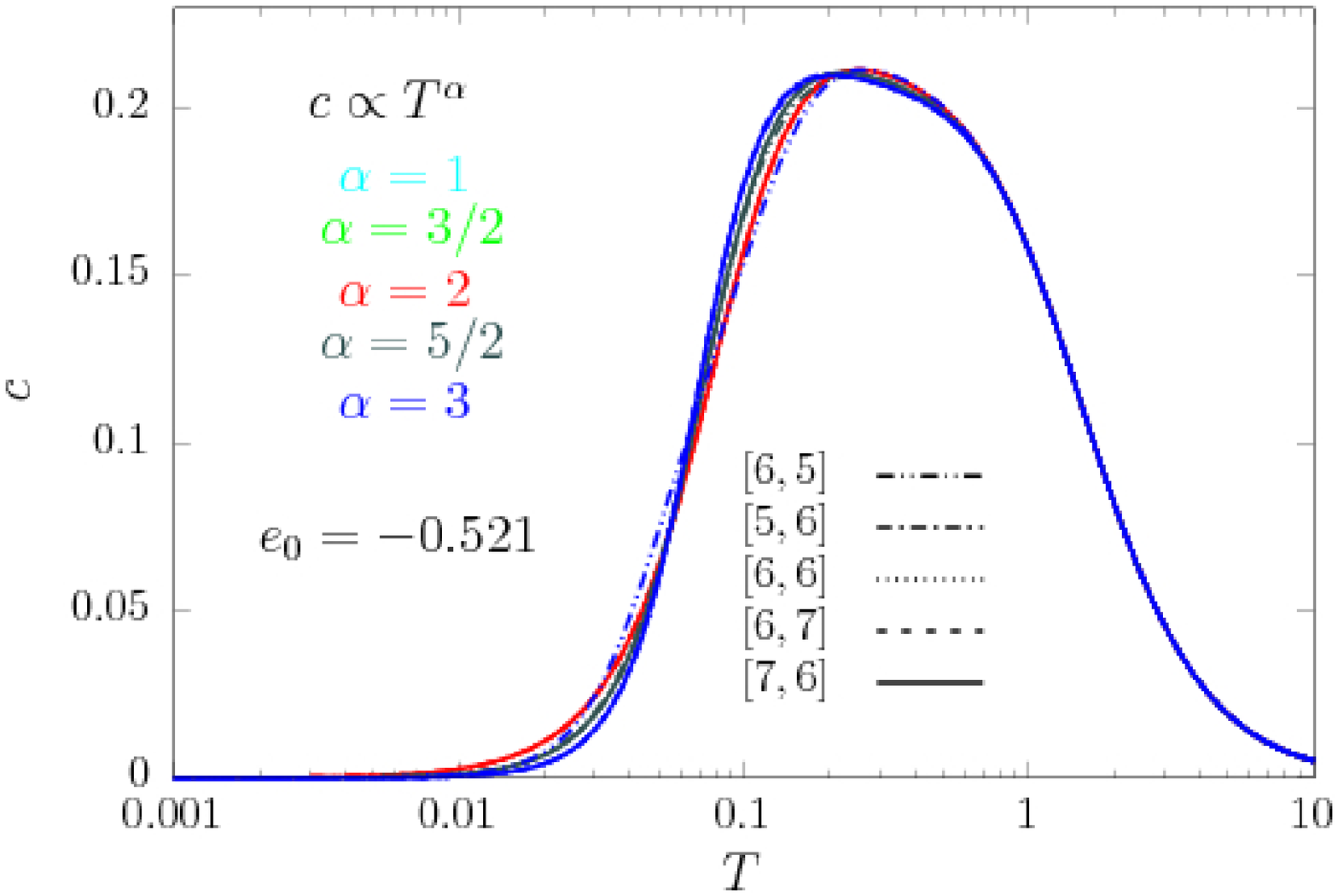} 
\protect
\caption{EM results for the specific heat;
gapless spectrum. 
Various $e_0=-0.521\ldots-0.515$ (from bottom to top) and $\alpha=1,3/2,2,5/2,3$ (cyan, green, red, magenta, blue, respectively).}
\label{fig20}
\end{figure}

Fig.~\ref{fig20} provides temperature profiles of the specific heat which complement those shown in Fig.~\ref{fig10} and Fig.~\ref{fig19}: 
We show $c(T)$ for $e_0=-0.521$, $e_0=-0.519$, $e_0=-0.517$, and $e_0=-0.515$ (from bottom to top)
and compare data for $\alpha=1$ (cyan), $\alpha=3/2$ (green), $\alpha=2$ (red), $\alpha=5/2$ (magenta), and $\alpha=3$ (blue).
The shown temperature profiles allow to estimate how close to each other are the various EM data based on different Pad\'{e} approximants.

\begin{figure}
\centering 
\includegraphics[clip=on,width=80mm,angle=0]{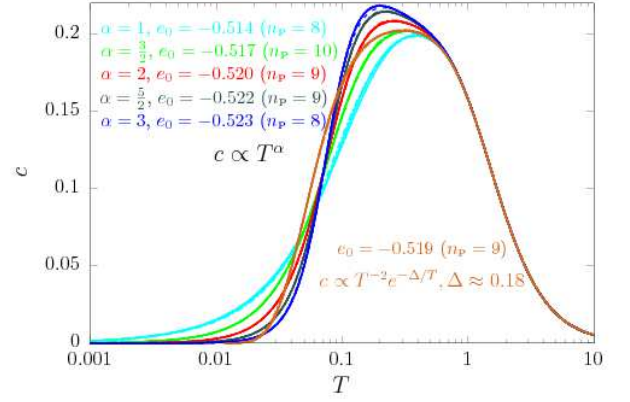} 
\protect
\caption{Specific heat $c(T)$ of the $S=1/2$ PHAF:
Comparison of EM data for gapped (light brown) and gapless (other colors) excitations.}
\label{fig21}
\end{figure}

In Fig.~\ref{fig21} we collect the best 
(i.e., for such a value of $e_0$ which gives the largest number of almost coinciding resulting curves) 
EM predictions for $c(T)$ for the gapped and the gapless spectrum.

\begin{figure}
\centering 
\includegraphics[clip=on,width=80mm,angle=0]{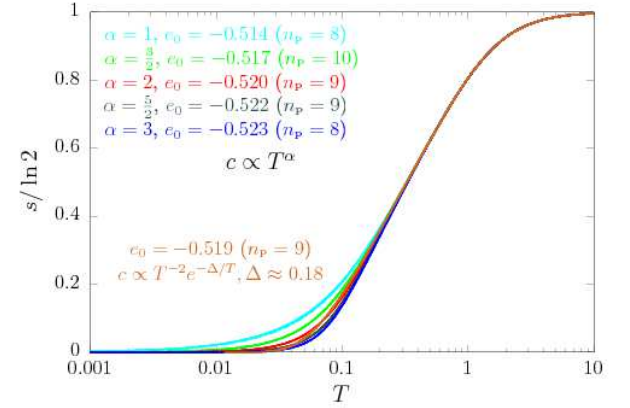} 
\protect
\caption{Comparison of EM data for the entropy for the gapped and gapless cases.}
\label{fig22}
\end{figure}

The best EM results for the temperature dependence of the entropy are shown in Fig.~\ref{fig22}.
Note, all curves for each color are indistinguishable in this figure.

\begin{figure}
\centering 
\includegraphics[clip=on,width=80mm,angle=0]{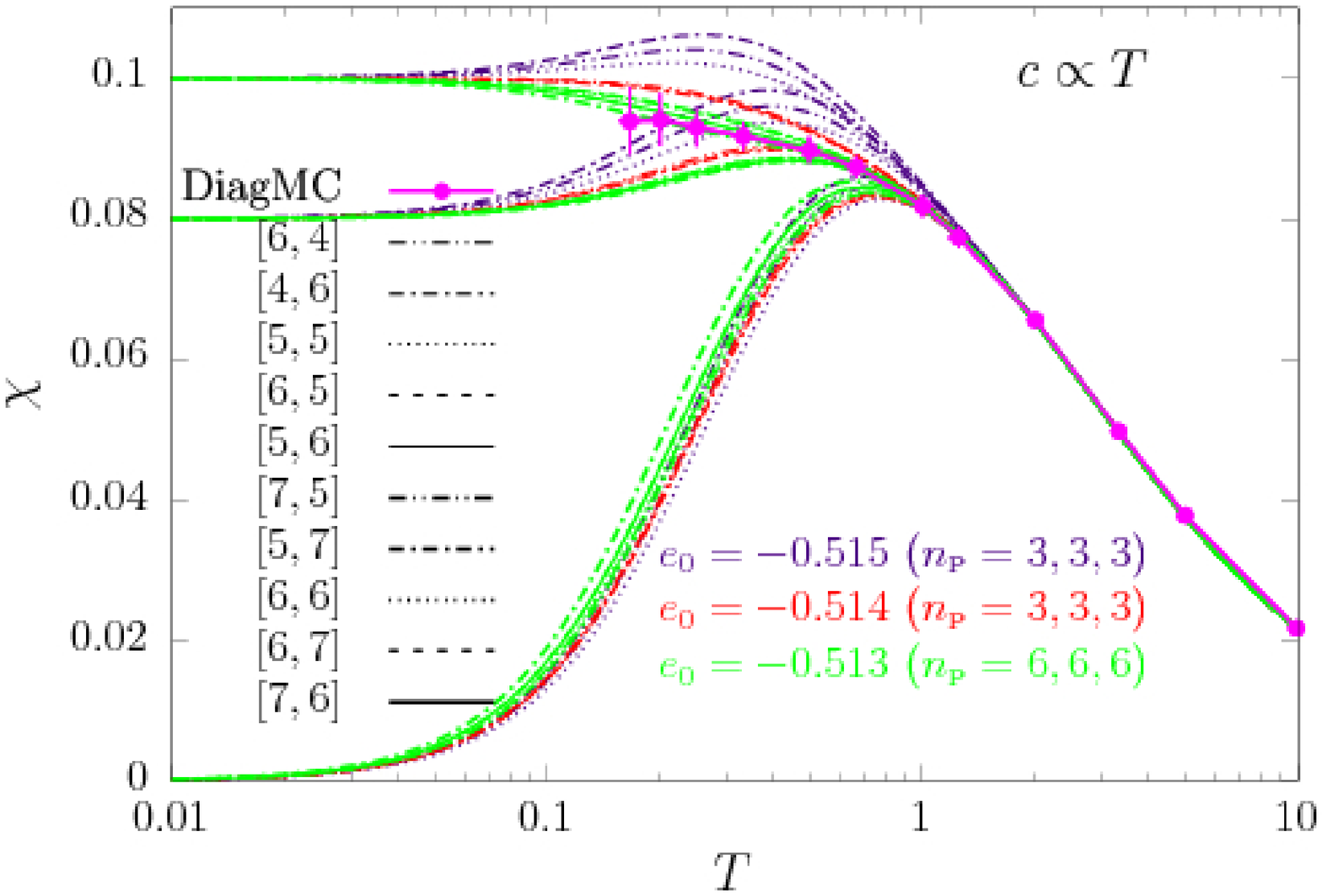} 
\includegraphics[clip=on,width=80mm,angle=0]{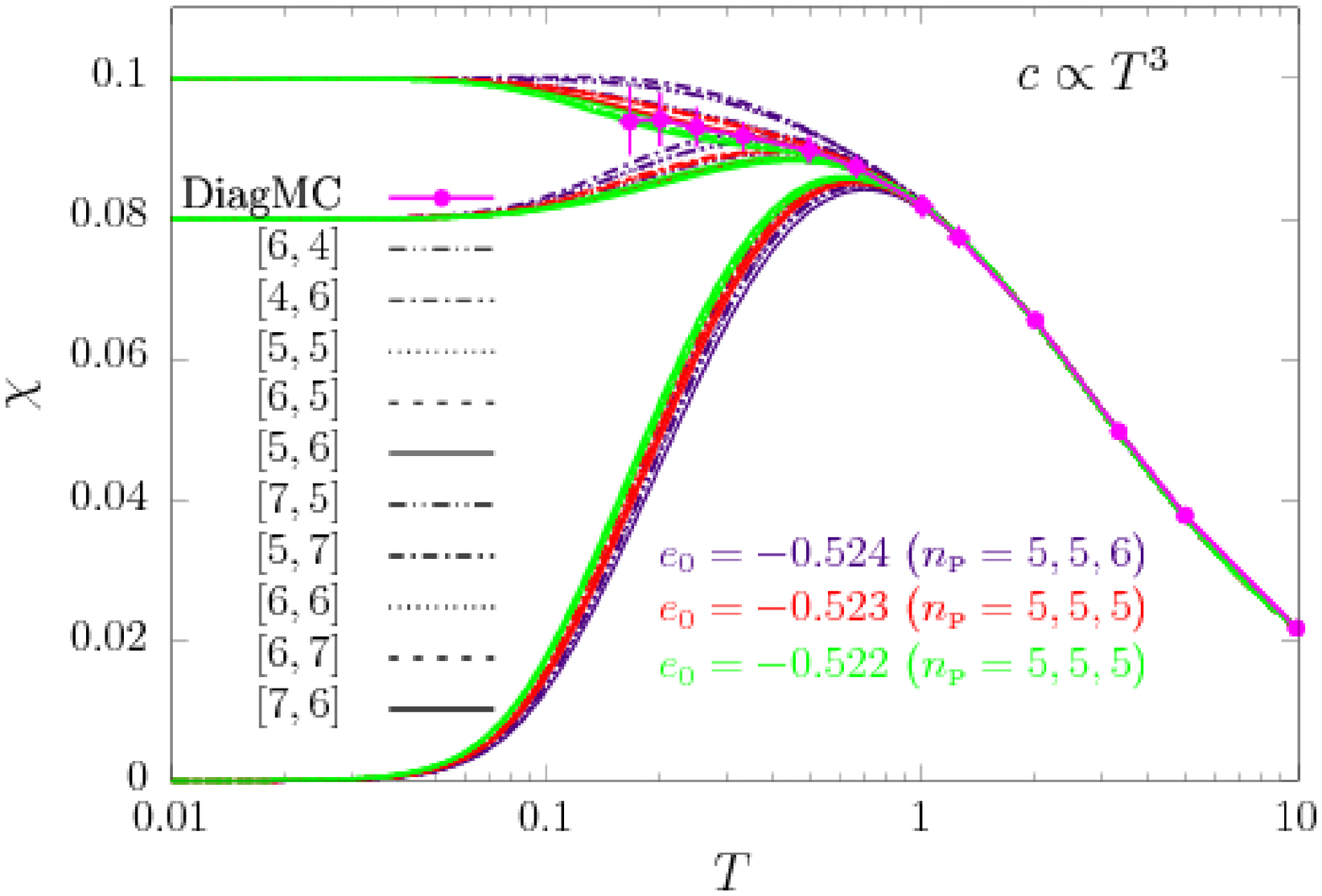} 
\protect
\caption{Supplement to Fig.~\ref{fig14};
EM results for the susceptibility;
gapless spectrum.
(Top) $e_0=-0.515\ldots -0.513$, $\alpha=1$.
(Bottom) $e_0=-0.524\ldots -0.522$, $\alpha=3$.
$\chi_0=0,0.08,0.1$.}
\label{fig23}
\end{figure}
 
Fig.~\ref{fig23} is supplementary to Fig.~\ref{fig14} of the main text.
It presents EM results for $\chi(T)$ for less favorable exponents $\alpha=1$ and $\alpha=3$.

\begin{figure}
\centering 
\includegraphics[clip=on,width=80mm,angle=0]{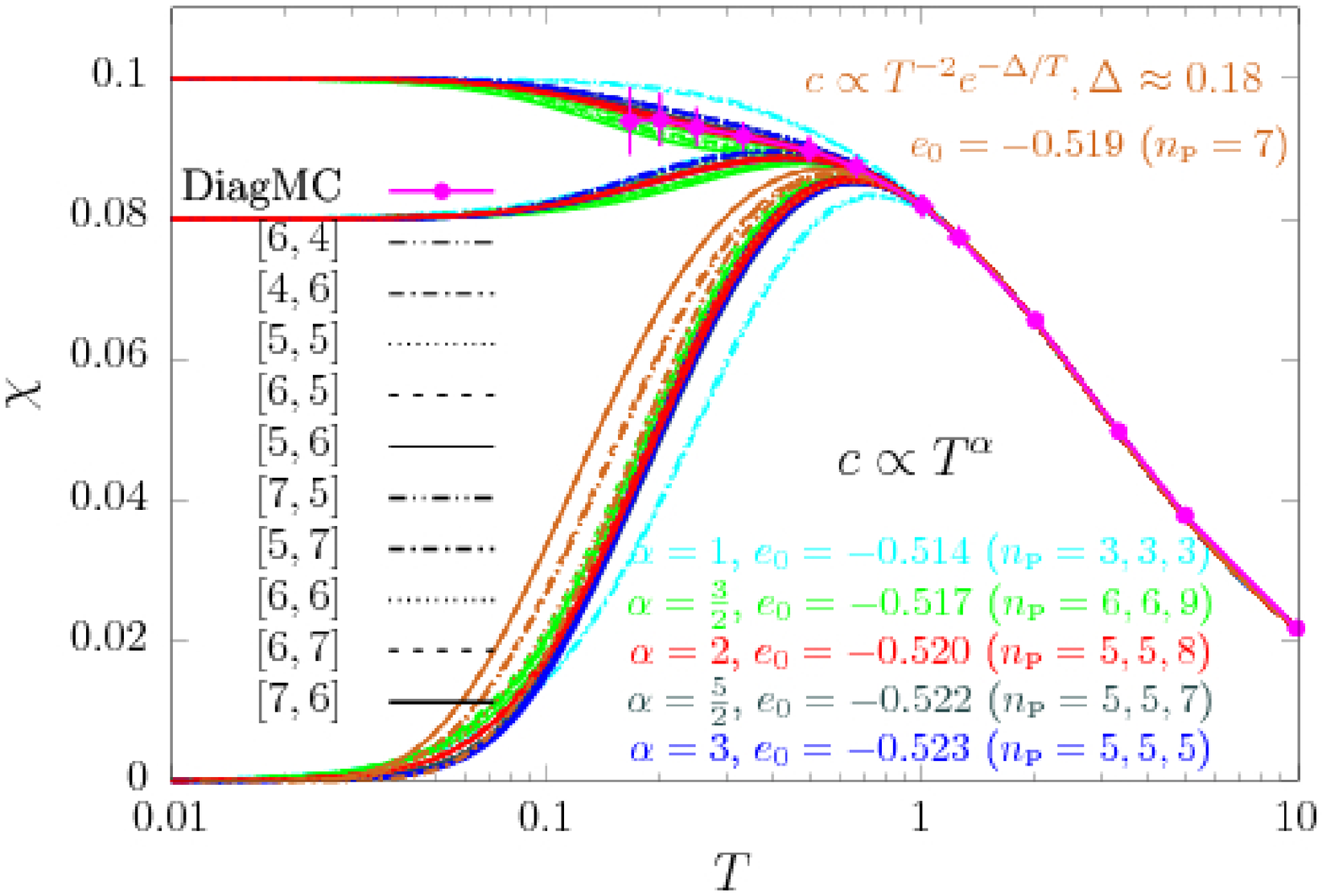}
\protect
\caption{Susceptibility $\chi(T)$ of the $S=1/2$ PHAF:
Comparison of EM data under the gapped (light brown) and gapless (other colors) assumptions.
$\chi_0=0,0.08,0.1$.}
\label{fig24}
\end{figure}

In Fig.~\ref{fig24} we collect the best 
(i.e., for such a value of $e_0$ which gives the largest number of almost coinciding resulting curves) 
EM predictions for $\chi(T)$ under the gapped assumption and the gapless assumption with several values of $\alpha$.
Note, 
$\chi(T)$ as it follows from the assumption about gapless singlet excitations but $\chi_0=0$ (all colors except light brown)
deviates stronger from the diagrammatic Monte Carlo result than $\chi(T)$ as it follows from the assumption about gapped excitations (light brown).
However, the agreement of different seven $\chi(T)$ curves for the latter case is rather poor.

\bibliography{phaf,js-own,js-mag}

\end{document}